\documentclass[format=acmsmall, screen=true]{acmart}

\usepackage{soul}
\usepackage{color}
\usepackage{booktabs} 
\usepackage{multirow}
\usepackage{subfig}
\usepackage{adjustbox}
\usepackage[utf8]{inputenc}

\acmJournal{CSUR}

\acmVolume{0}
\acmNumber{0}
\acmArticle{0}
\acmYear{0000}
\acmMonth{0}
\copyrightyear{0000}

\setcopyright{acmlicensed}

\acmDOI{0000001.0000001}

\received{February 0000}
\received[revised]{March 0000}
\received[accepted]{June 0000}

\begin{document}



\title[Integrated NFV/SDN Architectures: A Systematic Literature Review]{Integrated NFV/SDN Architectures: A Systematic Literature Review}  
\author{Michel S. Bonfim}
\orcid{1234-5678-9012-3456}
\author{Kelvin L. Dias}
\author{Stenio F. L. Fernandes}
\affiliation{%
	\institution{Universidade Federal de Pernambuco}
	\department{Centro de Informatica (CIn)}
	\city{Recife}
	\state{PE}
	\postcode{50740-560}
	\country{Brazil}
}

\begin{abstract}
  Network Functions Virtualization (NFV) and Software-Defined Networking (SDN) are new paradigms in the move towards open software and network hardware. While NFV aims to virtualize network functions and deploy them into general purpose hardware, SDN makes networks programmable by separating the control and data planes. NFV and SDN are complementary technologies capable of providing one network solution. SDN can provide connectivity between Virtual Network Functions (VNFs) in a flexible and automated way, whereas NFV can use SDN as part of a service function chain. There are many studies designing NFV/SDN architectures in different environments. Researchers have been trying to address reliability, performance, and scalability problems using different architectural designs. This Systematic Literature Review (SLR) focuses on integrated NFV/SDN architectures, with the following goals: i) to investigate and provide an in-depth review of the state-of-the-art of NFV/SDN architectures, ii) to synthesize their architectural designs, and iii) to identify areas for further improvements. Broadly, this SLR will encourage researchers to advance the current stage of development (i.e., the state-of-the-practice) of integrated NFV/SDN architectures, and shed some light on future research efforts and the challenges faced.
\end{abstract}

%
%

\ccsdesc[500]{General and reference~Surveys and overviews}
\ccsdesc[500]{Networks~Network design principles}
\ccsdesc[500]{Networks~Network services}

%
%

\keywords{Software-defined networking, network function Virtualization, network virtualization, cloud computing, mobile networks, resource provisioning, autonomic management, service-level agreement, quality of service, resource scheduling, resource management, scalability, elasticity, reliability, security
}

\thanks{This work is supported by grants from UFPE and CNPq (304422/2013-4, 305223/2016-0).

Authors’ addresses: M. Bonfim, K. Dias, and S. Fernandes (corresponding author), Centro de Informática (CIn), Universidade Federal de Pernambuco (UFPE), P.O.Box 7851 - Cidade Universitária, 50740-560, Recife - PE, Brazil; emails: {msb6, kld, stenio}@cin.ufpe.br.}

\maketitle

\renewcommand{\shortauthors}{Michel S. Bonfim et al.}

\section{INTRODUCTION}
\label{sec:int}

Proprietary network hardware equipment is everywhere in businesses, homes, and data center networks. Each vendor explores and exploits the maximum capability of its platforms as a way to meet the performance, reliability, and availability requirements demanded by the various types of users. However, such an approach has resulted in incompatibility between different manufacturer’s technologies. Restricted licensing agreements and proprietary source code have further contributed to this limitation.

Because of such incompatibility of platforms, as well as the need for network engineers to add new features into their networks (e.g., firewalls, load balancing), they often need to purchase new equipment from different vendors. Each item of equipment is responsible for a share of the traffic processing that requires specific management strategies. Difficulties in the management and configuration of such heterogeneous environments are the norm. The requirement to allow such flexibility in configuration and the deployment of new network functions must be met in new business and engineering models. The complexities of current networking environments result in high operational (OPEX) and capital (CAPEX) expenditure costs~\cite{white_nfv1}.

Network Functions Virtualization (NFV) aims at solving these problems by transferring networking functions from vendor-specific and proprietary hardware appliances to software hosted on Common-Off-The-Shelf (COTS) systems (a.k.a commodity platforms), i.e., with standard processing, memory, and storage components. These systems usually provide the network services in virtual machines (e.g., Virtual Network Functions - VNFs), each one performing different operations (e.g., firewall, packet inspection, routing, etc.)~\cite{white_nfv2}. NFV has the potential to allow cost reduction and the increase in speed of network expansion. Also, NFV has the potential to increase network flexibility for fast service delivery, an option difficult to achieve with traditional methods~\cite{Szabo2015}.

Software-Defined Networking (SDN) is a new network paradigm. Its main feature is the separation of the network control plane from the data plane, compared to current networks where the IP layer integrates both planes vertically into the network devices ~\cite{McKeown:2008:OEI:1355734.1355746}. In the SDN control plane, represented by a software called SDN Controller, which is responsible for decisions on how to handle the underlying network traffic concerning network policies and rules. The SDN Controller can run on COTS systems, separated from the forwarding devices. The data plane, deployed as network devices, is responsible for forwarding data according to a set of rules. The SDN controller allows the creation and management of such rules through an Application Programming Interface (API) in the Northbound interface. It does have direct control over the data plane elements through protocols in the Southbound interface. Such a separation provides some definite advantages, such as simplification and flexibility in network policy enforcement, facilitating network configuration, development, and fostering innovation~\cite{Kreutz201414}. It also brings research and development challenges that have attracted researchers from both industry and academia.

According to Mijumbi et al. (2016)~\cite{7243304}, ``NFV and SDN have a lot in common since they both advocate for a passage toward open software and network hardware''. Even with different purposes, NFV and SDN do indeed represent complementary paradigms and technologies capable of providing one consolidated solution. To this end, SDN can provide connectivity between VNFs in a flexible and automated way, thus simplifying network management. On the other hand, NFV can make use of SDN as part of a Service Function Chaining (SFC). In this case, both SDN Controllers and Management Applications can run as VNFs in a scalable environment and hence benefit from essential features, such as availability, reliability, and elasticity.

Some studies are tackling the integration of NFV and SDN in different environments (e.g., Cloud Computing, Wide Area Network, Customer Premise Equipment, 5G, etc.). Industrial and academic research studies address several challenges, such as reliability, overall performance, and scalability. Those studies use distinct architectural design rationale and functional and non-functional requirements. Although NFV/SDN architectures have clear potential benefits, they are still at an early stage of development. There are several open research questions to be answered and development issues to be addressed ~\cite{Kreutz201414,7243304,Batista2015}. 

It is worth emphasizing that there have been some initial efforts to review the body of knowledge on NFV and SDN, but most efforts treat them in isolation. Mijumbi et al. (2016) \cite{7243304} and Gil and Botero (2016) \cite{7437249} surveyed the state-of-the-art in NFV, whereas Kreutz et al. (2015) \cite{Kreutz201414} presented a survey exclusively on SDN. Furthermore, both Li and Chen (2015) \cite{7350211} and López et al. (2015) \cite{7272185} propose studies to integrate both technologies. However, there is still a need for a detailed vision of the different integrated NFV/SDN architectures (e.g., target environment, problems to solve, and architectural designs) as well as trends for research and development. We argue that the research community would benefit from an in-depth view of the NFV/SDN architectural designs so that researchers can have a clear picture of the past relevant studies as well as the current challenges. Therefore, our study fills an important gap by providing an in-depth view of the SDN/NFV architectures, as well as highlighting the challenges to further advances in this topic.

In this work, we cover the state-of-the-art of integrated NFV/SDN architectures. We aim at: i) investigating the characteristics (target environment and problems to solve) of integrated NFV/SDN solutions and current practices; ii) comparing their architectural designs (i.e., NFV framework design and tools, SDN APIs, and placement of SDN elements); and iii) identifying the challenges and the possibilities for improving them. To this end, we conducted a Systematic Literature Review (SLR) to provide an overview of this research area, based on a well-known methodological framework introduced by Kitchenham et al. (2009) \cite{Kitchenham20097}.  

SLR is an evidence-based approach used to identify, evaluate, and interpret all available evidence about a focused topic, in a repeatable and impartial manner ~\cite{kitchenham2004procedures}. For this, the SLR framework follows a predefined protocol with a set of steps to perform sources and studies selection and data extraction. In the end, results are synthesized from this well-defined approach by comparing the individual studies and providing consistent evidence of the research questions being posed.

Our original contributions are three-fold. First, this SLR provides an in-depth understanding of both state-of-the-art and state-of-the-practice of NFV/SDN architectural solutions, highlighting their main characteristics (e.g., potential deployment scenarios and problems raised) and their underlying architectural designs. Second, the SLR study identifies trends for future research and development as well as open research issues and challenges. Last, but not least, our SLR provides the necessary details for replicating it or broadening its scope in the future. 

The remainder of the article is organized as follows. Section \ref{sec:background} describes NFV and SDN technologies. Section \ref{sec:slr-plan} introduces the details of the adopted SLR. It explicitly defines the steps of the protocol and the strategies to retrieve the evidence, to allow this SLR to be reproduced and criticized by other professionals. Section \ref{sec:search} describes the search process that resulted from the SLR execution. Section \ref{sec:app-nfv-sdn} describes the target environments and problems addressed by the studies. Section \ref{sec:taxonomy} describes a taxonomy to organize the various decision-making levels for the design of NFV/SDN architectures. Such a taxonomy was derived by analyzing implementations found in the researched literature and reference architectures proposed by vendors and standardization bodies. Sections \ref{sec:nfvside} and \ref{sec:sdnside} provide the technical details of this taxonomy. Section \ref{sec:tools-nfv-sdn} presents the mainly auxiliary tools used in the studies to implement NFV/SDN architectures. Section \ref{sec:lessons} presents a brief description of the results obtained. Section \ref{sec:challenges} lists the challenges involved in developing NFV/SDN solutions. Section \ref{sec:threats} describes some threats to the validity of this study, to evaluate the quality of this research. Finally, Section \ref{sec:conclusions} concludes the article.
\section{BACKGROUND}
\label{sec:background}

\subsection{Network Functions Virtualization (NFV)} \label{sub:nfv}

NFV is transforming the computer and communication networks industry. NFV allows customers to transfer the networking functions from vendor-specific and proprietary hardware appliances to software hosted on COTS platforms  ~\cite{white_nfv1}. 

NFV provides the network services in virtual machines (VMs) working in Cloud infrastructures, where each VM performs different network operations (e.g., firewall, intrusion detection, Deep Packet Inspection, load balancing, etc.) \cite{white_nfv2}. Some benefits of deploying network services as virtual functions are ~\cite{white_nfv1}:

\begin{itemize}
\item Flexibility in the allocation of network functions in general-purpose hardware;
\item Rapid implementation and deployment of new network services;
\item Support of multiple versions of service and multi-tenancy scenarios;
\item Reduction in CAPEX costs by managing energy usage efficiently;
\item Automation of the operational processes, thus improving efficiency and reducing OPEX costs.
\end{itemize}

From 2012, the European Telecommunications Standards Institute (ETSI) has led the standardization process for NFV technology through the NFV Industry Specification Group (NFV ISG). The NFV ISG has already published tens of specifications documents, such as requirements, use cases, terminologies, proofs of concept, and the like \cite{ETSI2014}. These specifications allow researchers and engineers to have a clear picture of the elements of a particular NFV infrastructure.

\begin{figure}[htbp]
\centering
\includegraphics[scale=0.6]{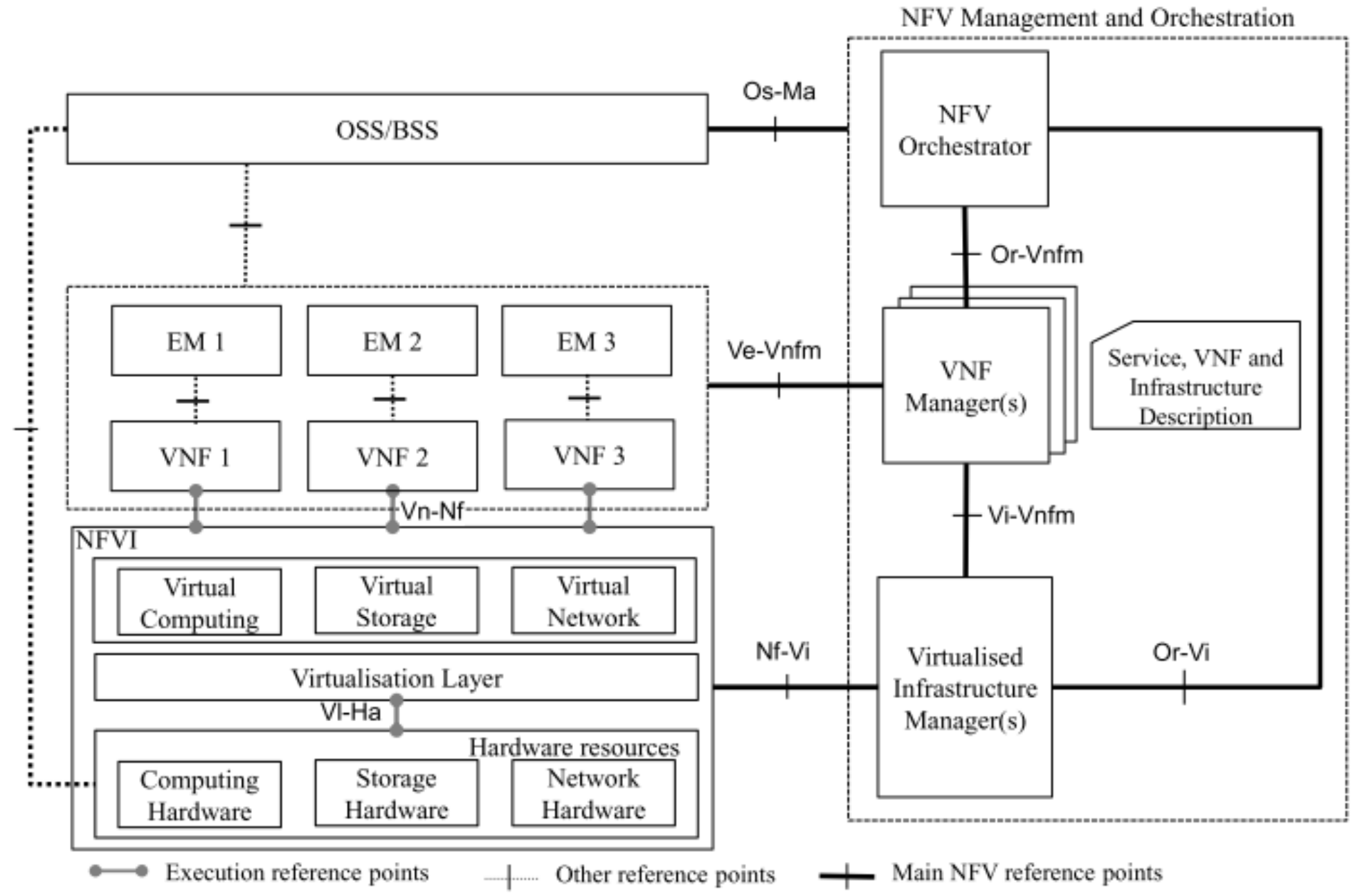}
\caption{NFV Architecture \cite{ETSI2014}.}
\label{fig:nfv-reference-architecture}
\end{figure}

Figure \ref{fig:nfv-reference-architecture} illustrates the high-level architecture for NFV, which comprises of three main functional blocks, as detailed below.

\begin{description}
\item[Virtual Network Functions (VNFs):] VNF is the virtualization of a certain network function, which should operate independently of the others. It may run on one or more virtual machines. A particular VNF can also be divided into several sub-functions called VNF Components (VNFCs). Elemental Management Systems (EMSs) can be used for VNF monitoring;
\item[NFV Infrastructure (NFVI):] NFVI comprises of all hardware and software required to deploy, operate, and monitor VNFs. To this end, NFVI has a virtualization layer necessary for abstracting the hardware resources (processing, storage, and network connectivity). It ensures the independence of the VNF software from the physical resources. The virtualization layer is usually composed of the server (e.g., Xen, KVM, VMware, etc.) and the network (e.g., VXLANs, NVGRE, OpenFlow, etc.) hypervisors. The NFVI Point of Presence (NFVI-PoP) defines a location for Network Function deployments as one or many VNFs.
\item[NFV Management and Orchestration (MANO):] MANO comprises three components: i) The Virtualized Infrastructure Manager (VIM), which  
manages and controls the interaction of VNFs with physical resources under its control (e.g., allocation, deallocation, and inventory); ii) the VNF Manager (VNFM), which is responsible for managing the VNF life-cycle (e.g., initialization, suspension, and termination); and iii) the NFV Orchestrator (NFVO), which is responsible for realizing network services on NFVI. It also performs monitoring operations of the NFVI as a way to collect information for operations and performance management.
\end{description}

Another component to be considered as part of the NFV framework is the Operations Support Systems and Business Support Systems (OSS/BSS). This element comprises the legacy management systems and assists MANO in the execution of network policies, either automatically or manually. 

\subsection{Software-Defined Networking (SDN)}

SDN is a new network paradigm that was designed to overcome the difficulty in developing and testing new solutions and protocols in production environments, where the underlying code running in business switches and routers are proprietary and closed \cite{McKeown:2008:OEI:1355734.1355746}.   

According to Kreutz et al. (2015), currently, both control and data planes are integrated into most commercial networking devices, which makes IP networks difficult to manage. Due to this, operators need to configure network policies into each device individually, often using low-level commands that are specific to the manufacturer. Further, automatic reconfiguration mechanisms, necessary for network adaptation during failures and load changes, are non-existent in today's networks. Such issues reduce the flexibility for deploying new network services and management strategies, as well as hindering development and innovation.

The main feature of the SDN paradigm is the separation of the control and data planes. It has clear advantages where network programmability is achieved through the centralization of the control plane in conjunction with the availability of open APIs, thus making easier the process of creating and deploying new network applications. SDN provides simplification and flexibility in network policy enforcement, facilitating network configuration and management \cite{Kreutz201414}.  

The control plane, represented by a software called the SDN Controller, is responsible for decisions on how to handle network traffic, assuming the role of the ``brain'' of the network. The SDN Controller can run on COTS platforms, separated from the network equipment. The data plane, represented by the network devices, is responsible for forwarding traffic according to a set of rules \cite{Kreutz201414}. Such rules are created at and managed by the SDN Controller. The SDN controller has a global view of the network topology and has direct control over the data plane elements through a southbound protocol, such as OpenFlow \cite{onfspecopenflow13} (detailed in Section \ref{sub:openflow}).

\begin{figure}[ht]
\centering
\includegraphics[scale=0.3]{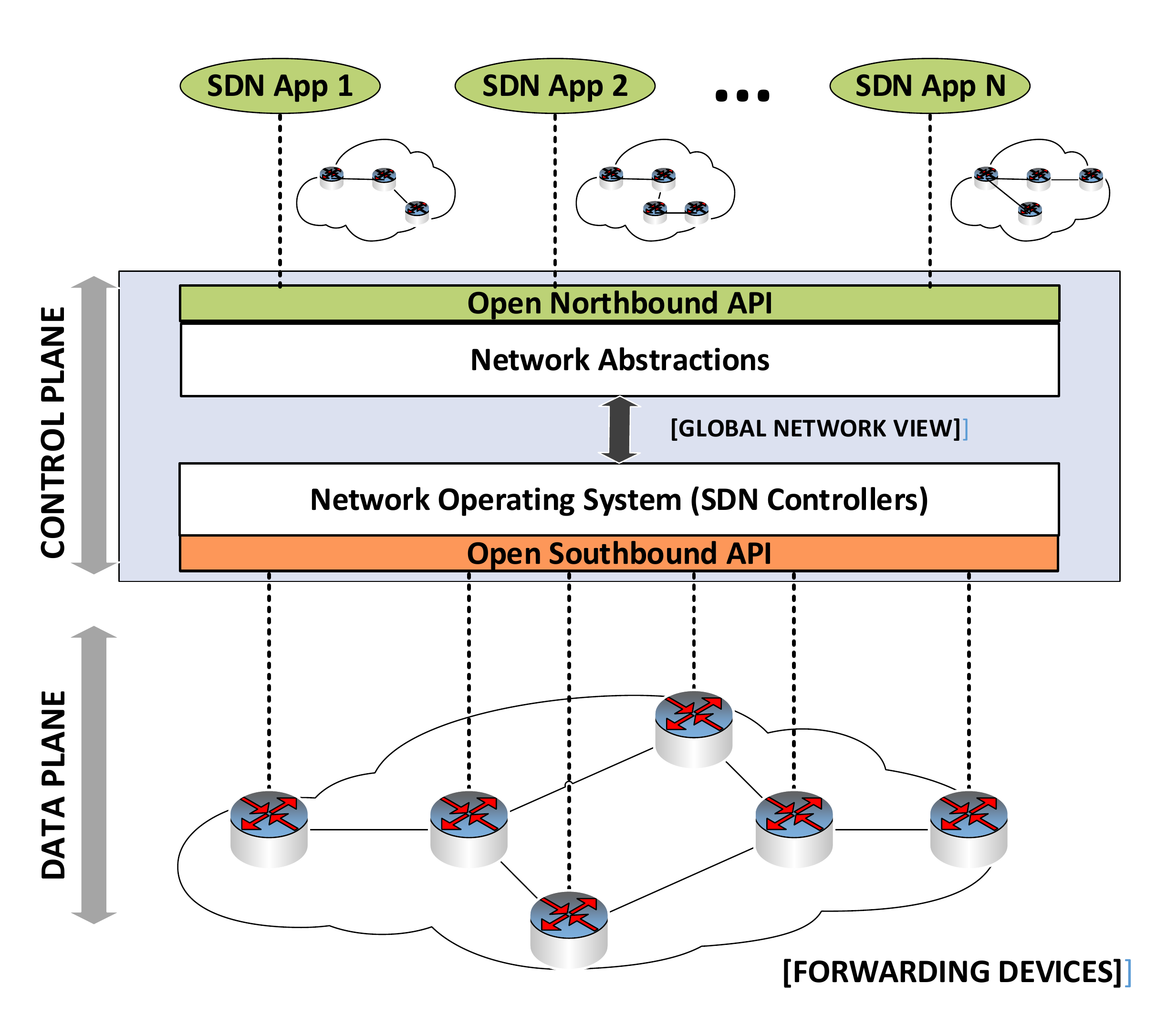}
\caption{SDN Reference Architecture \cite{McKeown:2008:OEI:1355734.1355746,Kreutz201414,onfspecopenflow13}.}
\label{fig:sdn-reference-architecture}
\end{figure}

Figure \ref{fig:sdn-reference-architecture} shows the given three layers of an SDN architecture and the APIs responsible for the interaction between them. The SDN Northbound API is responsible for providing support for communication between the application layer and the control plane layer. It also includes support for SDN Applications, such as traffic engineering, routing, firewall, quality of service, etc. The Southbound API is responsible for the communication between the SDN Controllers and switches. 

\section{SYSTEMATIC LITERATURE REVIEW PLANNING}
\label{sec:slr-plan}

This section presents the adopted plan to perform the SLR. This phase aims to define the way the review is executed, including the research questions and the procedure for sources and studies selection. 

\subsection{Research Questions} \label{subsec:slr-plan-rq}

To identify the state-of-the-art of integrated NFV/SDN solutions and open issues, we initially focus our study on the following research questions.

\begin{itemize}

\item \textbf{Q1) In which environments are the integrated NFV/SDN solutions applied? }

This question aims at mapping the actual environments (e.g., Enterprise Networks, WANs, CPEs, Data Centers, Wireless Networks, etc.) in which the proposed integrated NFV/SDN solutions have been tested and deployed.

\item \textbf{Q2) What are the problems that such integrated NFV/SDN solutions are trying to solve?}

This question aims at identifying the issues (e.g., middlebox and network virtualization, vCPE, reliability, scalability, dynamic service function chaining, performance, etc.) that NFV/SDN solutions are trying to tackle. We also aim to classify them according to their respective target environments.

\item \textbf{Q3) What are the differences among the design architectures of integrated NFV/SDN solutions?}

This question seeks clarification on how the studies have been proposing design architectures of integrated NFV/SDN solutions. Answering this question requires a classification of the proposed architectures using a subset of their characteristics extracted from the ETSI documentation \cite{etsi_sdn_in_nfv}, published in December 2015. These characteristics include NFV framework design and tools, SDN Northbound and Southbound APIs, the placement of the SDN elements in the NFV Framework, the use of multiple SDN controllers, and the like. We aim at identifying the differences among such 
designs based on the types of target environments and general objectives, including their main advantages and disadvantages. 


\end{itemize}

\begin{table}[ht]
\caption{List of Terms and Synonyms.}
\label{tab:keywords}
\begin{minipage}{\columnwidth}
\begin{center}
\begin{tabular}{lll}
\toprule
& \textbf{Group 1} & \textbf{Group 2} \\ \midrule
Term 1 & Software-Defined Networking & Network Function Virtualisation \\ 
Term 2 & Software Defined Networking & Network Function Virtualization \\ 
Term 3 & Software Defined Network & NFV \\
Term 4 & Software-Defined Network & \\ 
Term 5 & SDN & \\ 
\bottomrule
\end{tabular}
\end{center}
\end{minipage}
\end{table}%

\subsection{Sources Selection}

To find the relevant evidence to answer the research questions, a set of sources must be selected to perform the search of primary studies. We now describe the criteria used to select such sources, the search strings, and the sources identification. 

For the selection criteria of sources, we considered the availability of articles on the Web and the existence of advanced search mechanisms using keyword and filters based on content type (conference publications, journals, and magazines, etc.) and year of publication. We considered only studies in the English language. Therefore, we selected the following web search engines: ACM Digital Library, Engineering Village, IEEE Xplore, Science Direct, Scopus, Springer Links, and Web of Science.

To compose our search string, we considered the keywords listed in Table \ref{tab:keywords}, where each group represents a keyword with its synonyms. The general form of the search string is shown as follows:

\begin{description}
	\item[Search String:] (([G1,T1] \textbf{OR} [G1,T2] \textbf{OR} [G1,T3] \textbf{OR} [G1,T4] \textbf{OR} [G1,T5]) \textbf{AND} ([G2,T1] \textbf{OR} [G2,T2] \textbf{OR} [G2,T3]))
\end{description}

The literature searches were performed manually using all the selected web search engines. Table \ref{tab:searchstrings} shows the composition of the search string per web search engine. Regarding ACM Digital Library's search queries, the symbol ``+'' replaces the logical operator ``AND'' while the space symbol replaces the logical operator ``OR''. In Web of Science's search queries, the symbol ``TS'' determines that search is limited to the following fields within a record: Title, Abstract, Author Keyword, Keywords Plus.

\begin{table}[ht]
\caption{List of Search Strings per Web Search Engine.}
\label{tab:searchstrings}
\begin{minipage}{\columnwidth}
\begin{center}
\begin{tabular}{p{5cm}p{8cm}}
\toprule
\textbf{Web Search Engine} & \textbf{Search String} \\ \midrule
\textbf{ACM Digital Library} & +(``Software-Defined Networking'' ``Software Defined Networking'' ``Software Defined Network'' ``Software-Defined Network'' ``SDN'') +(``Network Function Virtualisation'' ``Network Function Virtualization'' ``NFV'') \\ 

\textbf{Engineering Village, IEEE Xplore, Science Direct, Scopus, and Springer Links} & (``Software-Defined Networking'' OR ``Software Defined Networking'' OR ``Software Defined Network'' OR ``Software-Defined Network'' OR ``SDN'') AND (``Network Function Virtualisation'' OR ``Network Function Virtualization'' OR ``NFV'')\\ 

\textbf{Web of Science} & TS=((``Software-Defined Networking'' OR ``Software Defined Networking'' OR ``Software Defined Network'' OR ``Software-Defined Network'' OR ``SDN'') AND (``Network Function Virtualisation'' OR ``Network Function Virtualization'' OR ``NFV''))\\ 
\bottomrule
\end{tabular}
\end{center}
\end{minipage}
\end{table}

\subsection{Procedure for Studies Selection} \label{subsec:slr-plan-pss}

A priori, all studies in the English language obtained from web search engines were selected as primary studies. These primary studies then went through a studies selection and evaluation process, based on three stages. One researcher (M. Bonfim) was assigned to evaluate the selected studies. An article is included for further processing in the next steps when it is approved in the previous one. Otherwise, the article is discarded. 

Below, we describe the three stages of the studies selection and evaluation process:

\begin{itemize}

\item \textbf{Stage 1:} Eliminate studies selected as primary studies based on exclusion criteria. An article will be only included in the following stages if it proposes an integrated NFV/SDN solution. This stage considers only information provided in abstract and conclusion.

\item \textbf{Stage 2:} Eliminate studies selected in Stage 1 based on exclusion criteria. An article will be only included for the following stages if it proposes an integrated NFV/SDN solution and describes its architecture design. This stage evaluates all content of the articles.

\item \textbf{Stage 3:} In this stage, studies selected at Stage 2 pass for a quality screening. An article will be excluded if it does not meet the following quality criteria:

\begin{itemize}
\item \textbf{QC1:} Is there a clear statement of the goals (i.e., target environments and problems to solve) of the research?
\item \textbf{QC2:} Is the architecture design well detailed? In other words, is it possible identify the used tools, the place of SDN elements and NFV framework design? 
\item \textbf{QC3:} Are the experiments realized to evaluate the ideas presented in the study?
\end{itemize}

\end{itemize}

Each criterion has three possible responses: Yes, Partly, or No. ``Yes'' responses count for 1 (one) point, ``Partly'' count for 0.5 points and ``No'' count for 0 (zero) points. To be accepted, a paper must obtain a score of greater or equal to 2 (two) as described in equation \ref{eq:quality}:

\begin{equation}
  QC1 + QC2 + QC3 \geq 2.0
  \label{eq:quality}
\end{equation}

Finally, at the end of execution, we included some reports from Proof of Concepts (PoCs) registered in ETSI NFV ISG PoC Projects \footnote{http://www.etsi.org/technologies-clusters/technologies/nfv/nfv-poc}, regarding NFV/SDN solutions provided by different vendors and carrier networks. 
\section{SEARCH RESULTS}
\label{sec:search}

The initial search was performed in April 2016. Initially, a total of 1644 articles were identified (Identity Phase) as primary studies. In the Identity Phase, 907 duplicate findings were removed from the result set. It is important to emphasize that we do not delimit a specific range of years for the searching process. Then we started the execution of the three stages of selection, as described in Subsection \ref{subsec:slr-plan-pss}.

In Stage 1 (Screening Phase), having reviewed all abstracts and conclusions, we considered only articles that proposed an integrated NFV/SDN solution. In this case, we selected 138 studies and discarded 769. In Stage 2 (Screening Phase), we considered only records included in Stage 1. After evaluating all the content of articles, we considered only those that described the design of the integrated NFV/SDN solution. In this case, we selected 88 studies and discarded 50. In Stage 3 (Eligibility Phase), studies selected in Stage 2 were passed for a quality screening, described in Subsection \ref{subsec:slr-plan-pss}. In this stage, we discarded 40 studies that did not meet the quality criteria, with the result that 48 studies were included (Included Phase) for data extraction.

At the end of the execution process, twelve (12) Proof of Concepts (PoCs) reports (registered in ETSI NFV ISG PoC Projects \footnote{http://www.etsi.org/technologies-clusters/technologies/nfv/nfv-poc}) are included, generating a total of 60 studies for data extraction. The PoCs are part of the Hot Topic 01\footnote{http://nfvwiki.etsi.org/index.php?title=HT01\_-\_Use\_of\_SDN\_in\_an\_NFV\_architectural\_framework} - ``Use of SDN in an NFV architectural framework'', which includes SDN/NFV solutions provided by different vendors and carrier networks. Table \ref{tab:pocsincluded} lists these PoCs.

\begin{table}[ht]
\caption{List of Proof of Concepts (PoCs) included in the execution process.}
\label{tab:pocsincluded}
\begin{minipage}{\columnwidth}
\begin{center}
\begin{tabular}{cp{11cm}}
\toprule
\textbf{Identifier} & \textbf{Title} \\ \midrule
POC\#1 & CloudNFV - Open NFV Framework Project. \cite{poc_etsi_1} \\ 
POC\#2 & Service Chaining for NW function selection in Carrier Networks. \cite{poc_etsi_2} \\ 
POC\#8 & Automated Network Orchestration. \cite{poc_etsi_8} \\ 
POC\#13 & Multi-Layered Traffic Steering for Gi-LAN. \cite{poc_etsi_13} \\ 
POC\#16 & NFVIaaS with Secure SDN-controlled WAN Gateway. \cite{poc_etsi_16} \\ 
POC\#21 & Network intensive and compute intensive hardware acceleration. \cite{poc_etsi_21} \\ 
POC\#23 & E2E orchestration of Virtualised LTE Core-Network functions. \cite{poc_etsi_23} \\ 
POC\#26 & Virtual EPC with SDN functions in Mobile Backhaul Networks. \cite{poc_etsi_26} \\ 
POC\#27 & VoLTE Service based on vEPC and vIMS architecture. \cite{poc_etsi_27} \\ 
POC\#28 & SDN Controlled VNF Forwarding graph. \cite{poc_etsi_28} \\ 
POC\#34 & SDN-enabled Virtual EPC Gateway. \cite{poc_etsi_34} \\ 
POC\#38 & Full layer-7 stack fulfillment, activation, and orchestration of VNFs in carrier networks. \cite{poc_etsi_38} \\ 
\bottomrule
\end{tabular}
\end{center}
\end{minipage}
\end{table}

The following step is the data collection for each selected work in the Included Phase. One (1) researcher (M. Bonfim) performed the data extraction, extracting the following properties from each study: (i) author's identification; (ii) article type (conference paper, journal article, report, etc.); (iii) publication description (title, ISSN, date, DOI, etc.); (iv) work's title and abstract; (v) environments in which the NFV/SDN solution is applied; (vi) problems that NFV/SDN solution try to solve; (vii) technical aspects related to NFV/SDN solution proposed, and (viii) quality criteria evaluation.

After performing the SLR, during the article's revision, another 14 relevant references were found and/or recommended by experts in the field to complement the SLR findings. The total number of research papers is now 74, including new references from 2016 until 2017. These papers respected the same Exclusion (Stage 2), and Quality Criteria (Stage 3) adopted previously. All results and discussions presented in this article were derived from these 74 studies.  

There is a growing interest from both the academia and the industry in integrating NFV and SDN technologies. The number of studies increased from 1 research paper in 2013 to 43 in 2015, and this number has been rising since then. The researchers prefer scientific conferences to publish their studies (35 articles), followed by journals (26 articles) and PoC reports (12 articles).

The reader can find more details regarding the data collection (extracting documents) at the GitHub link\footnote{Link for data collection archives: https://github.com/michelsb/SLRNFVSDNFiles.git}.
\section{APPLICATION FOR NFV/SDN ARCHITECTURES}
\label{sec:app-nfv-sdn}

This section aims at answering the first and second research questions, thus relating the different environments where the NFV/SDN architectures found are applied, in addition to identifying the main problems those architectures are trying to solve. 

First, it is necessary to present the definitions for the following terms, which will eventually appear in the studies.

\begin{description}
	\item[Reliability:] Relates to fault tolerance, disaster recovery, and full isolation;
	\item[Elasticity in Network Function:] It is the possibility to scale network services dynamically at runtime in an automated fashion \cite{Szabo2015};
	\item[Flexibility:] Service providers can design the layout of service chains without considering the physical network \cite{ding_openscaas:_2015};
	\item[Security:] It means privacy, authentication, or authorization;
	\item[Scalability:] Service providers can augment the number of service chains without worrying about constraining flow tables \cite{ding_openscaas:_2015};
	\item[Dynamic Service Chaining:] Applying different policy-based traffic steering to flows in a certain SFC;
	\item[QoS Management:] Trying to optimize the use of network capacity through QoS techniques.
\end{description}

\subsection{Unifying Computer and Network Resources} \label{sub:unify} 

This concept aims at creating an abstraction layer (AL) on both the computational and network resources as a way to provide a unique and centralized view of the whole environment. The NFV/SDN solutions initially focused on creating this Abstraction Layer (AL) to delivery intelligent network services - regarding performance and reliability - for different customer profiles such as end, retail, and enterprise users, as well as Over The Top (OTT) providers, and developers \cite{etsi_sdn_in_nfv}. 

\begin{figure}[htbp]
	\centering
	\begin{adjustbox}{width=0.5\textwidth}
		\includegraphics[]{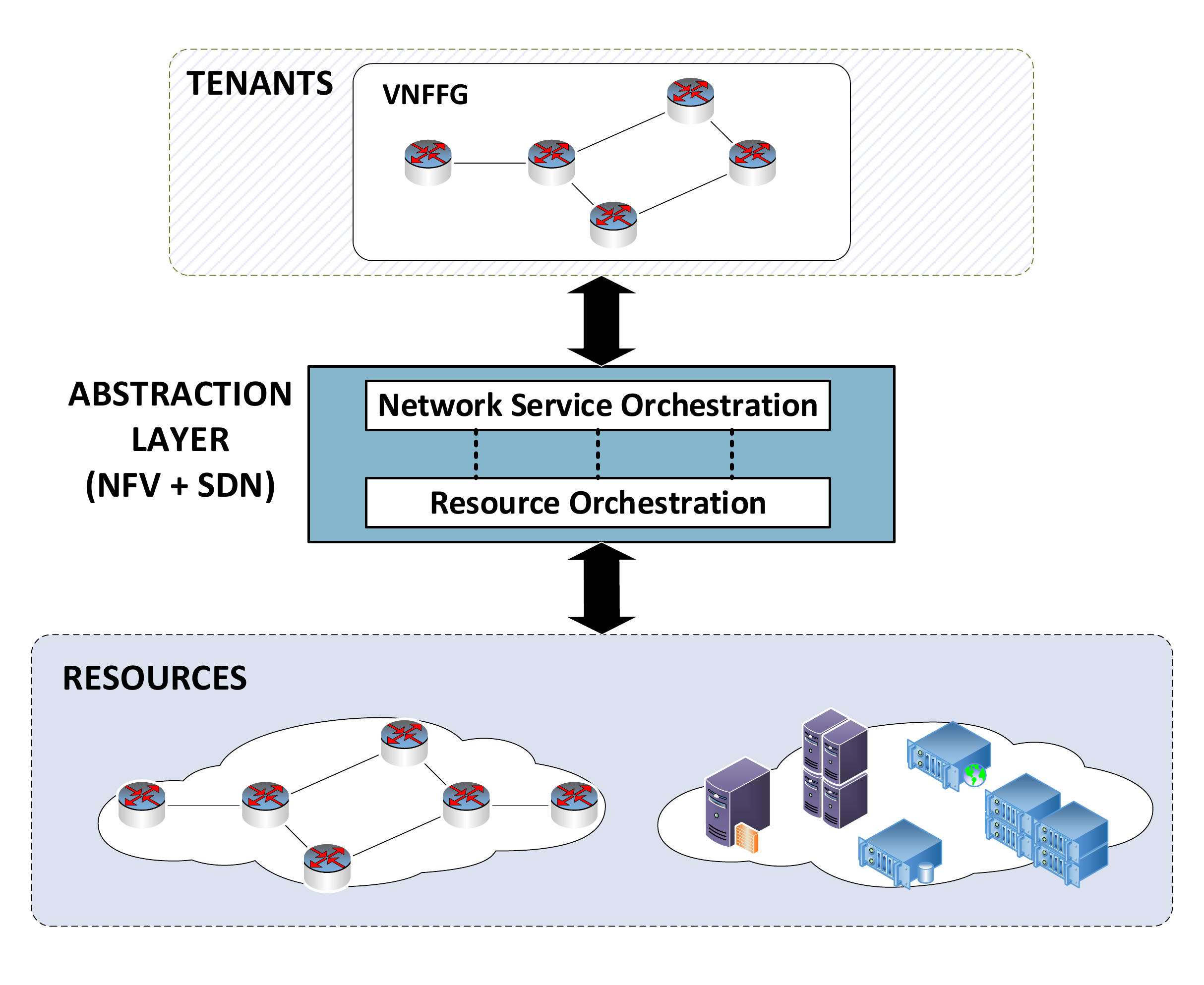}
	\end{adjustbox}
	\caption{Abstraction layer for unifying computer and network resources.}
	\label{fig:abstract}
\end{figure}

The AL is responsible for two orchestration functions (see Figure \ref{fig:abstract}), namely Resource Orchestration (RO) and Network Service Orchestration (NSO) \cite{etsi_nfv_man}. RO utilizes NFV VIM component and SDN to perform a global resource management with resource virtualization provisioning and managing. On the other hand, the NFVO uses the NSO functions to implement the life-cycle management of Network Services (VNFFGs), coordinating groups of VNF instances as network services. These functions use the services exposed by the VNFM and by the RO allowing joint instantiation and configuration along with connectivity and dynamic changes management. 

In this context, different architectures try to achieve this unification, such as UNIFY \cite{6809448,6984052} and T-NOVA \cite{tnova}, both EU-funded 7\textsuperscript{th} Framework Programme (FP7) projects.

T-NOVA aims at implementing an SDN-based MANO framework to manage a federated network and cloud resources. Its primary objective is to deliver third-party network functions (NFs) to operator's customers in an automated and optimized manner, introducing a ``Network Function Store''. For NF management, an Orchestrator platform was developed on top of common open source components such as OpenStack \cite{openstack} and OpenDaylight \cite{opendaylight}. Besides, the WICM (WAN Infrastructure and Connectivity Manager) provides network connectivity between NFVI-PoPs (Points of Presence) and manages traffic steering in virtual networks. 

\begin{figure}[htbp]
	\centering
	\begin{adjustbox}{width=0.7\textwidth}
		\includegraphics[]{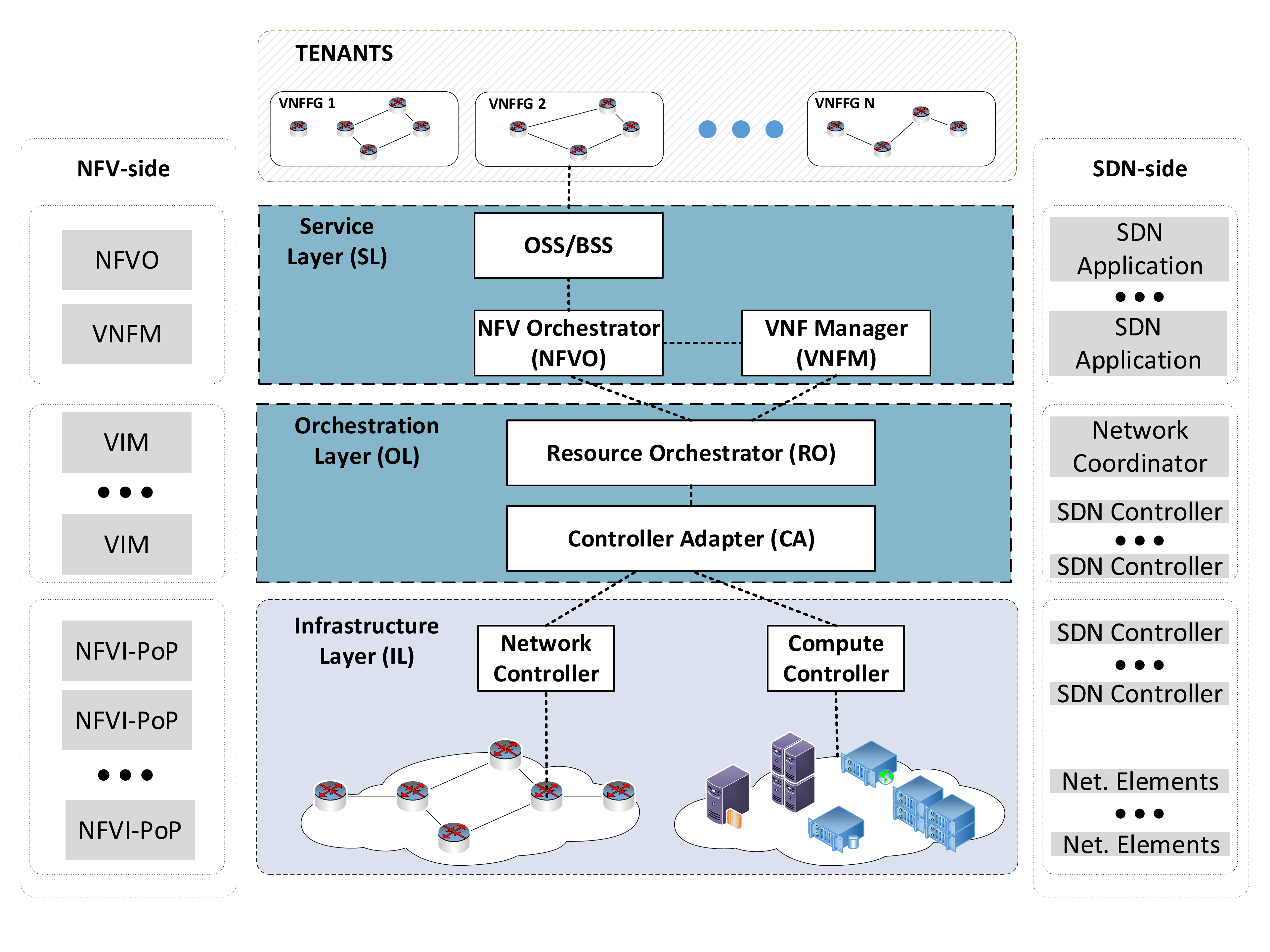}
	\end{adjustbox}
	\caption{UNIFY architecture.}
	\label{fig:unify}
\end{figure}

On the other hand, UNIFY seeks to unite computer and network resources in a common management framework. The UNIFY NFV/SDN architecture aims at creating and managing the dynamic end-to-end network services from the home and enterprise networks to the operator's data center. It provides a MANO framework that integrates both Cloud and WAN domains and includes three layers, namely the Service Layer (SL), the Orchestration Layer (OL), and the Infrastructure Layer (IL). Figure \ref{fig:unify} shows our simplified view of the UNIFY architecture, highlighting the main functional components of ETSI's NFV (left box) and ONF's SDN (right box) reference models.   

The SL comprises business management concerned with service life-cycle, providing Operation Support System (OSS) and Business Support System  (BSS) functions related to services from different tenants (e.g., enterprise users, service providers, etc.). The SL also executes the NFVO and VNFM functions for the NS and VNF life-cycle management. It is noteworthy that the SL management functions are infrastructure-agnostic, dealing only with the offered services \cite{sonkoly_unifying_2015}.

The OL acts as a VIM component - providing resource orchestration (RO module) to deliver virtual resources views to SL - and policy enforcement. It also includes the Controller Adapter (CA), and a multi-domain, multi-technology, and multi-vendor controller. The CA provides computing and networking abstraction by collecting virtualized resources from lower layer domain-specific controllers and organizing them into a global virtualized resource view. The CA offers an independent technology control to the RO module. Finally, in the OL, SDN is used for the creation and integration of virtual networks in both domains.    

The IL manages all IT and network resources (physical and virtual) needed for the VNF execution. For this, it uses two types of domain-specific controllers, the Compute Controller to manage computational resources, and the Network Controller to manage network resources. The IL considers different kinds of resources such as SDN enabled network nodes (e.g., OpenFlow switches), and cloud-enabled data centers (e.g., OpenStack, CloudStack, etc.).

The UNIFY architecture has been used as a basis for the implementation of several NFV/SDN solutions to different problems, such as Middleboxes Virtualization and virtualized Customers Premises Equipment (vCPE) (see Sections \ref{sub:midvirt} and \ref{sub:vcpe}).

\subsection{On-demand and Application-specific Traffic Steering}

Traffic Steering is the ability to direct users' requests to the appropriate service/content sources. Traffic steering might be based on many factors such as the available networking resources and capabilities on the client and server side, user's permissions and location, and the like. For example, a certain user may request a video streaming service that has stringent application performance requirements. In this case, on-demand and application-specific traffic steering could guarantee efficient network resource usage and better Quality of Experience (QoE) for the user.

In the NFV Framework context, SDN can enhance traffic steering between VNFs, providing dynamic service chaining. With the separation of control and data planes, SDN enables the exchange of information between the application and network layers, allowing users' services to have an overview of the general state of the network, and to make intelligent decisions (to meet service requirements) on how to steer traffic through VNFs better.

\begin{figure}[htbp]
	\centering
	\begin{adjustbox}{width=0.5\textwidth}
		\includegraphics[]{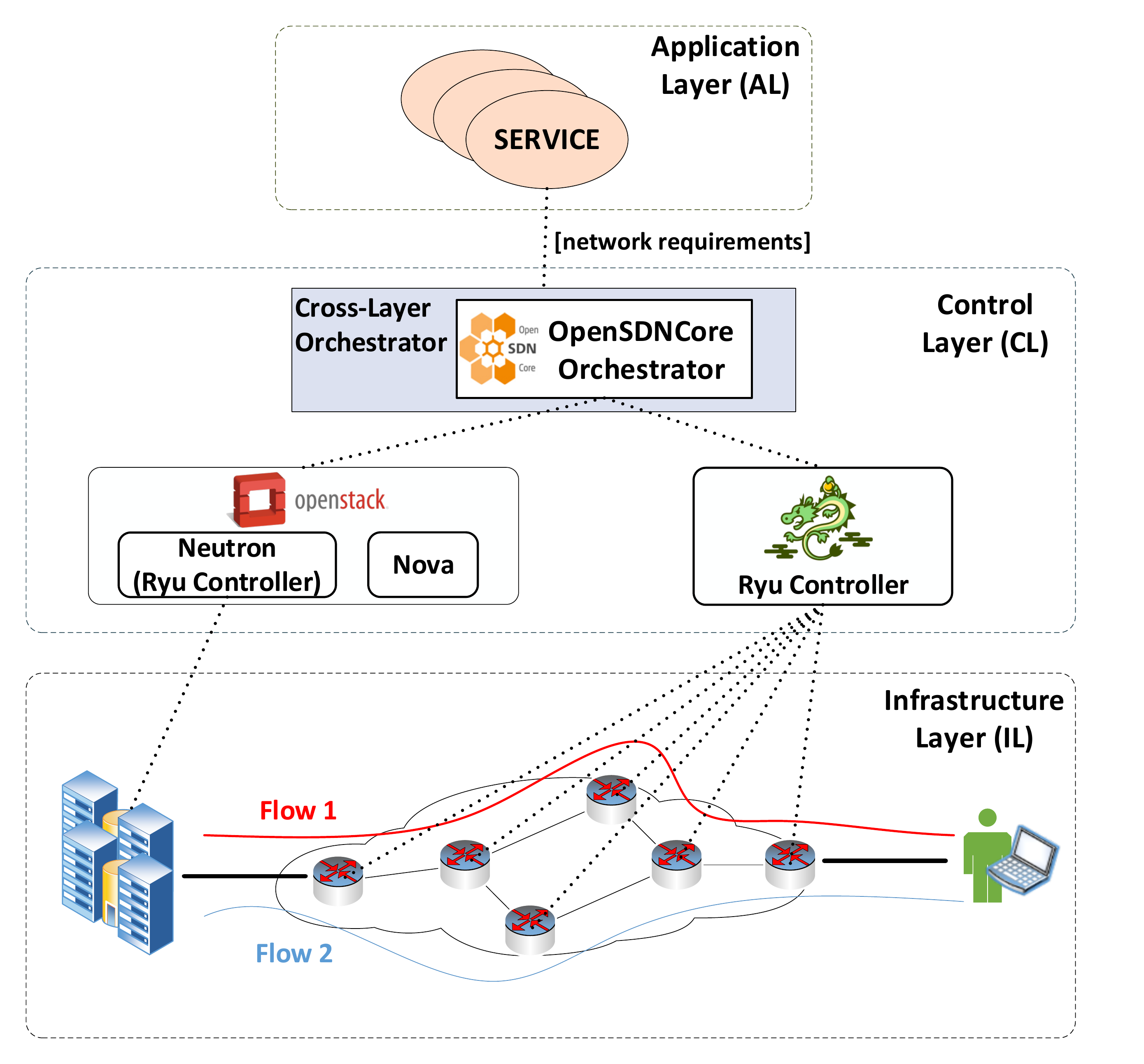}
	\end{adjustbox}
	\caption{NFV/SDN architecture for application-specific traffic steering.}
	\label{fig:carella}
\end{figure}

Carella et al. (2015) \cite{carella_cross-layer_2015} proposed an NFV/SDN architecture to provide a cross-layer interface between the application and network layers, to allow the deployment of network services with on-demand and application-specific traffic steering. Figure \ref{fig:carella} shows our detailed view of this architecture. It comprises three layers: Application, Control, and Infrastructure. The Application Layer consists of network services. These services must interface with the Control Plane API to communicate their network requirements (e.g., bandwidth, maximum latency). The Control Plane has a global view of computer and network resources and provides the traffic steering capabilities to the Application Layer. Its main component is the Cross-Layer Orchestrator (CLO) that acts as an NFVO and VNFM to manage the lifecycle of services over the cloud (OpenStack-based) and WAN domains (OpenFlow-based). The CLO was implemented over the OpenSDNCore Orchestrator \cite{opensdncore}, using Java programming language.

\subsection{Middleboxes Virtualization}\label{sub:midvirt}

According to \cite{Sherry:2012:MMS:2342356.2342359}, middleboxes have been increasingly used in enterprise networks (45\% of network devices). They are deployed to increase performance (e.g., traffic shaping, load balancing, and TCP optimization) and to provide security functionalities (e.g., firewalls, Intrusion Detection and Prevention systems - IDPS, and Deep Packet Inspection - DPI) for both incoming and outcoming traffic .

However, hardware-based middleboxes have the following disadvantages \cite{cziva_container-based_2015}. First, they incur high operational costs (OPEX) due to the management complexity. These middleboxes come from different manufacturers and must be deployed, configured, and managed individually. Second, hardware-based middleboxes incur capital costs (CAPEX). When new network functions are necessary, enterprises must purchase one or more middleboxes due to the inflexibility in proprietary hardware that creates vendor lock-in and limits innovation.

NFV/SDN architectures can be used to deal with these challenges. The primary objectives are to reduce both CAPEX and OPEX and to provide fast delivery of network function, elasticity, and dynamic service chaining. In these works, NFV manages virtual middleboxes, while SDN provides interconnection between VNFs to delivery Service Function Chaining (network services).

\begin{figure}[htbp]
\centering
\subfloat[Bounce Redirection.]{\includegraphics[width=0.5\textwidth]{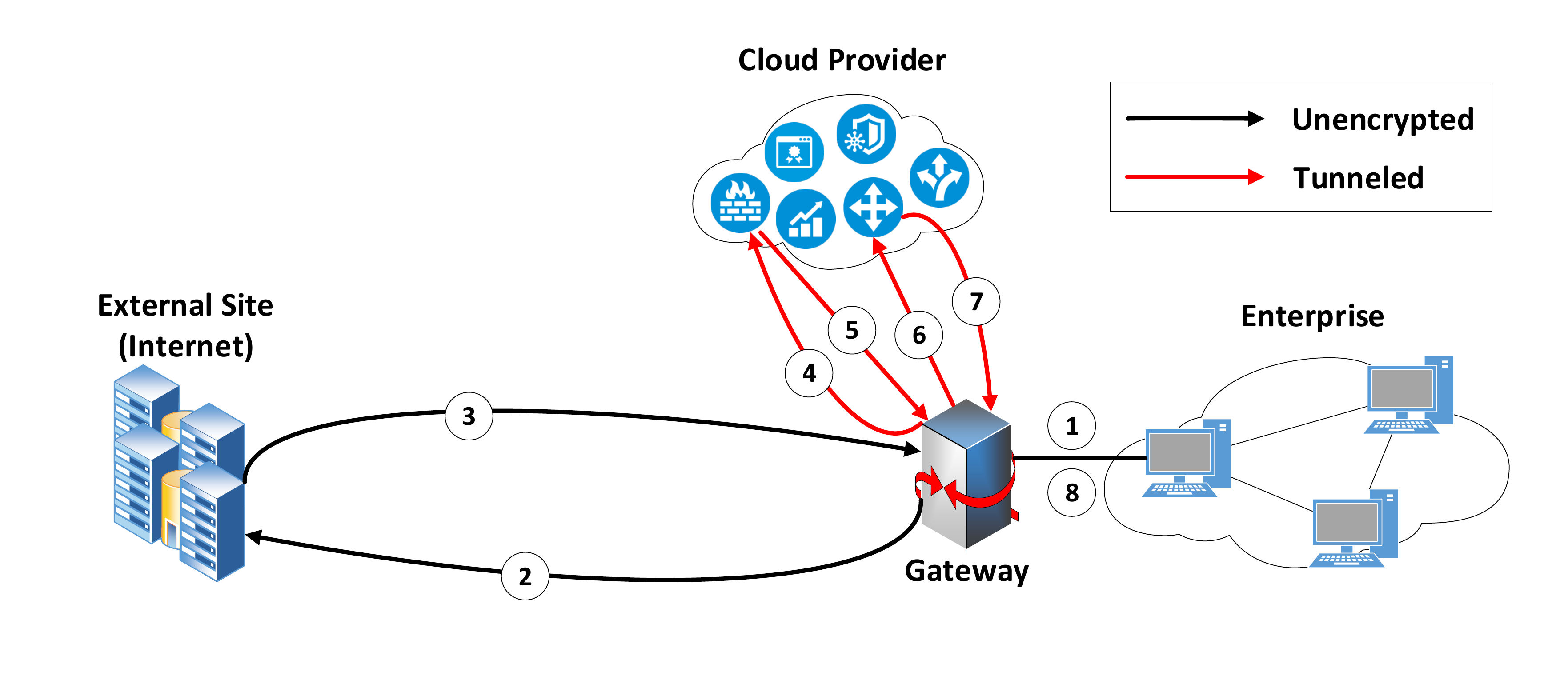}%
\label{fig:mid-virt1}}
\subfloat[IP Redirection.]{\includegraphics[width=0.5\textwidth]{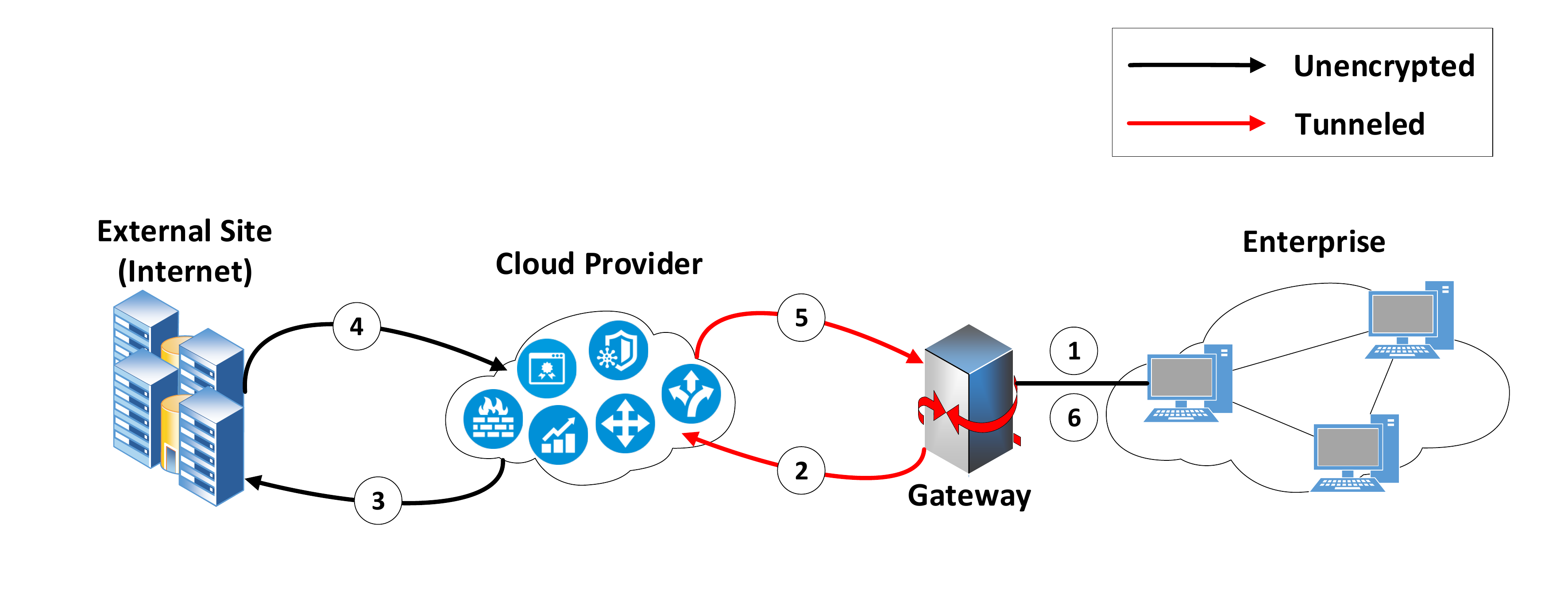}%
\label{fig:mid-virt2}}
\caption{Middleboxes virtualization with NFV and SDN \cite{Sherry:2012:MMS:2342356.2342359}.}
\label{fig:mid-virt}
\end{figure}

In this context, the authors in \cite{Sherry:2012:MMS:2342356.2342359} consider two approaches (see Figure \ref{fig:mid-virt}) for redirecting the traffic to virtualized middleboxes for further processing, namely Bounce and IP redirections. In the case of Bounce Redirections, a certain enterprise gateway uses tunneling techniques to redirect both ingress and egress traffic to the virtual middleboxes (grouped into service chains), as shown in Figure \ref{fig:mid-virt1}. It requires minimal configuration (i.e., few static rules) at the gateway since it only redirects traffic to a cloud provider hosting the middleboxes. However, this approach might increase the end-to-end delay due to these redirections for each packet. On the other hand, to avoid the extra round-trips of the Bounce Redirection (see Figure \ref{fig:mid-virt1}), IP Redirection allows routing traffic directly to/from the cloud provider, as shown in Figure \ref{fig:mid-virt2}. In this approach, the cloud will be located in the middle of the communication between enterprise and external sites. However, the provider must announce naming and addressing on the company's behalf (e.g., DNS redirection).

NFV/SDN architectures \cite{poc_etsi_23,batalle_implementation_2013,schulz-zander_opensdwn:_2015,lin_extended_2015,sonkoly_unifying_2015,deng_vnguard:_2015,cziva_gnfc:_2015,cziva_container-based_2015,rossem_deploying_2015,callegati_implementing_2015,7945849} have been proposed to deal with Middleboxes Virtualization. Some of them will be presented below.

\begin{figure}[htbp]
	\centering
	\begin{adjustbox}{width=0.5\textwidth}
		\includegraphics[]{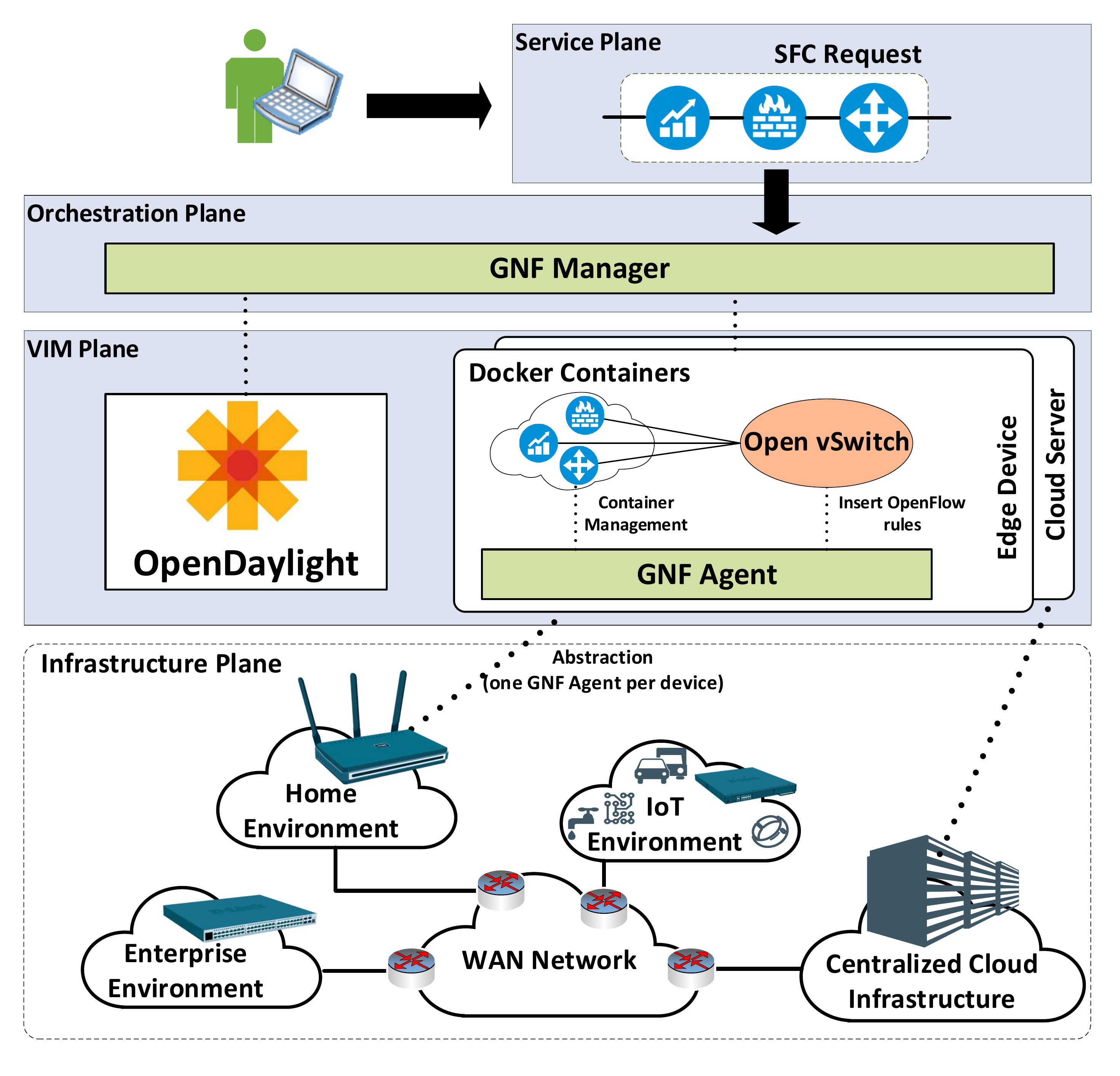}
	\end{adjustbox}
	\caption{The GNF platform \cite{7945849}.}
	\label{fig:cziva}
\end{figure}

The works of Cziva et al. (2015) proposed an NFV/SDN framework\footnote{https://netlab.dcs.gla.ac.uk/projects/glasgow-network-functions}, so-called Glasgow Network Functions (GNF), to deploy and manage container-based network services in public \cite{cziva_gnfc:_2015} and private \cite{cziva_container-based_2015,7945849} Cloud environments. This framework aims at overcoming the limited network reconfigurability in these scenarios, delivering network programmability and fast deployment of new network services. As shown in Figure \ref{fig:cziva}, GNF is composed of 4 planes: the infrastructure encompasses all the physical resources of network and computations, where we have only NFV Centralized Cloud Infrastructures in \cite{cziva_container-based_2015}, and incorporated with edge devices (e.g., CPEs, virtual routers, and IoT gateways) in \cite{7945849}. The VIM and Orchestration planes are responsible for Resource Orchestration. For this, the operator must deploy the GNF Agent on all Cloud servers and all edge devices. This feature has two functions: i) local VNF instantiation by using Docker Engine \cite{docker} for fast deployments and low resource utilization, and ii) local traffic steering management by using OpenFlow rules (via OVSDB) and virtual switches. Also, GNF uses the OpenDaylight controller for network connectivity in the NFVI. The GNF Manager is responsible for receiving NFV service requests (Service plane) and performing the necessary operations using the OpenDaylight and GNF Agent instances. 

In addition, Sonkoly et al. (2015) \cite{sonkoly_unifying_2015} extended the UNIFY architecture (see Section \ref{sub:unify}) to create a service function chaining control plane. This solution aims to support SFC in distributed cloud scenarios, where VNFs from the same SFC can run in different NFVI-PoPs. A prototype framework, called Extensible Service Chain Prototyping  Environment (ESCAPE), was implemented in Python on top of a POX Controller. This prototype works with two domains in the Infrastructure Layer (IL): Cloud and OpenFlow. The OpenStack Cloud Platform \cite{openstack} and the OpenDaylight Controller \cite{opendaylight} perform the management of Cloud domains while VNFs are deployed as KVM virtual machines (running a Click process). The OpenFlow domains handle transport networks with Linux nodes running Open vSwitches (OpenFlow support). The POX Controller \cite{pox} (network management) and the NETCONF/YANG (VNF management) manage these domains while VNFs are deployed as distinct processes (Linux cgroups) and run network functions implemented in Click Modular Router.

Finally, Deng et al. (2015) \cite{deng_vnguard:_2015} proposed the VNGuard framework that uses NFV to provide fast and dynamic virtual firewalls in a cloud environment, for the protection of Virtual Networks. Aimed at considering VN's changeable topology (VMs dispersion and migration), this framework uses SDN to provide fast and flexible traffic steering to virtual firewalls. They used OpenStack as a VIM and ClickOS for VNF development. CloudLab\footnote{https://www.cloudlab.us/} is a testbed that provides an Infrastructure as a Service (IaaS) for cloud-based experiments. 

\subsection{Virtualized Customers Premises Equipment (vCPE)}\label{sub:vcpe}

CPE (Customers Premises Equipment) means any equipment (router, modem, etc.) within the customer domain that receives a communication service. CPEs have been a barrier to the current goals of both telecommunications companies and service providers, due to the high cost of maintenance, management difficulties, and the impossibility of remote upgrades. As an alternative, a solution is the CPE virtualization using an NFV architecture, also known as Virtualized CPE (vCPE) \cite{etsi_nfv_use_cases}.

\begin{figure}[htbp]
	\centering
	\begin{adjustbox}{width=0.7\textwidth}
		\includegraphics[]{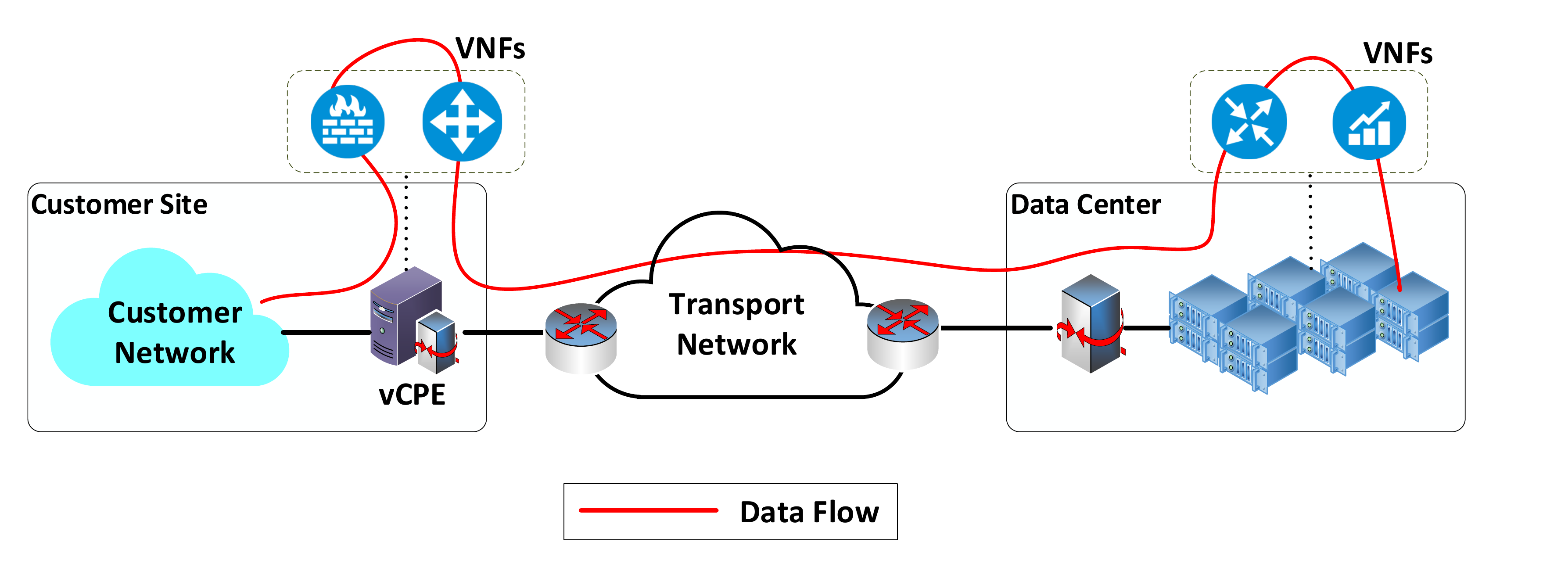}
	\end{adjustbox}
	\caption{NFV/SDN architecture for Virtualized Customers Premises Equipment (vCPE).}
	\label{fig:vcpe}
\end{figure}

According to an IHS Markit Survey\footnote{NFV Strategies Service Provider Survey - https://technology.ihs.com/572348/nfv-strategies-service-provider-survey-2016}, published in 2016, 100\% of consulted service providers said they intend to deploy NFV at some point. 81\% expect to roll out this deployment by 2017. Most service providers (more than 80\%) have a preference for deploying vCPE.

vCPE is a service in which some or all of the functions associated with CPE are virtualized ~\cite{etsi_nfv_use_cases}. One of the main problems related to CPEs virtualization is how to instantiate network services in distributed infrastructures (using multiple NFVI-PoPs) \cite{cerrato_toward_2015}. In this type of scenario (see Figure \ref{fig:vcpe}), also called Distributed NFV \cite{Gittik2014}, the VNFs are placed either in the service provider Cloud platform (Cloud CPE) or the on-premise CPE, depending on where they are most efficient regarding latency, available resources, etc. SDN has been the technology adopted to implement the communication management of different scenarios (e.g., Cloud, CPE, and WAN) to provide Distributed NFV.

\begin{figure}[htbp]
	\centering
	\begin{adjustbox}{width=0.5\textwidth}
		\includegraphics[]{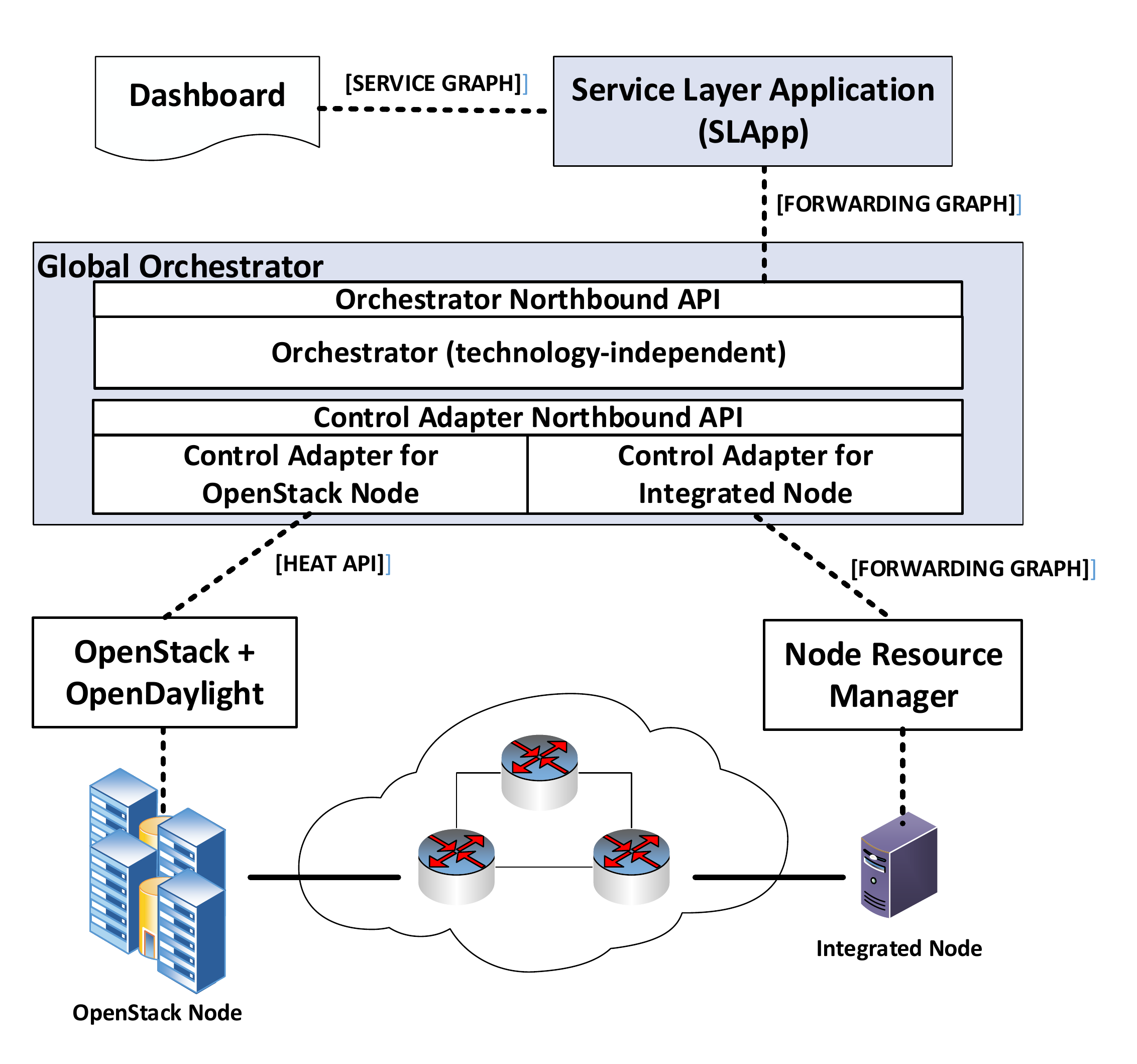}
	\end{adjustbox}
	\caption{An NFV/SDN architecture design for vCPE.}
	\label{fig:cerrato}
\end{figure}

Cerrato et al. (2015) \cite{cerrato_toward_2015} proposed a service-oriented NFV/SDN architecture for Telco networks that delivers generic network services selected by telecom operators (DHCP and NAT) or end users (BitTorrent client). The deployment of these network services can occur in a distributed manner either in the telecom data center or the CPE. Figure \ref{fig:cerrato} shows our simplified view of this architecture.  

This solution is based on the UNIFY architecture (see Section \ref{sub:unify}), including its three layers. The Service Layer Application (SLApp) represents the UNIFY SL enabling different players (operators and end users) to select their network services. For this, the SLApp includes an authentication mechanism and provides a high-level data model for defining flexible network services (including traffic steering primitives), called Service Graph (SG). 

Also, the Global Orchestrator (GO) represents the UNIFY Orchestration Layer (OL). The GO manipulates the Forwarding Graph (FG) received from SLApp to enable the network service deployment according to the VNF requirements and infrastructure capabilities. To allow distributed NFV, the GO implements multiple Control Adaptors to coordinate different infrastructures and an Orchestrator component responsible for the centralized coordination of multiple Control Adaptors. Then, the GO selects one of the infrastructures to implement all the network service requested. The authors proposed two different infrastructures (UNIFY Infrastructure Layer - IL) to host network services: the integrated node and the OpenStack node. 

The integrated node represents the CPE (home gateway). It receives an FG from the GO through the Node Resource Manager (NRM) via REST API. The NRM will instantiate all VNFs using Docker containers, DPDK process or any hypervisor supported by \textit{libvirt}. For traffic steering, the NRM uses an extensible Data-Path daemon (xDPd) to create an OpenFlow switch (and its correspondent Controller) for each FG. Separately, the OpenStack node represents the telco data center and uses the OpenStack Cloud Platform for network service deployment. In this case, the KVM hypervisor creates the VNFs, and the OpenDaylight and Open vSwitch control the traffic steering. 

The works of Soares \cite{soares_cloud4nfv:_2014,soares_toward_2015} proposed the Cloud4NFV platform, an NFV/SDN framework for Telco network virtualization. This platform considers multiple NFVI-PoPs and WAN domains when deploying new Service Function Chaining. Cloud4NFV considers a topology with multiples customer sites (NFVI-PoPs). All NFVI-PoPs include an OpenStack distribution working as a Cloud VIM and an OpenDaylight controller to provide VNF connectivity. The VNFs are CPE functions, and they are deployed as VMs in the NFVI-PoP closest to the customer. Further, Cloud4NFV includes a WAN VIM to provide a view of a unified WAN domain connecting the NFVI-PoPs and Telco Data Center. 

\subsection{Wireless Networks}

Due to the increasing popularity of wireless networks, new requirements have arisen, such as mobility support, programmability, fast delivery of network services, performance, and security \cite{schulz-zander_opensdwn:_2015}. However, the management and configuration of today's large WiFi networks are complex and inflexible, ignoring the application requirements or user needs. We present the problems addressed by NFV/SDN architectures designed for different wireless networks scenarios.

\subsubsection{\textbf{Wireless LAN}} \label{sub:wlan}

\begin{figure}[htbp]
	\centering
	\begin{adjustbox}{width=0.5\textwidth}
		\includegraphics[]{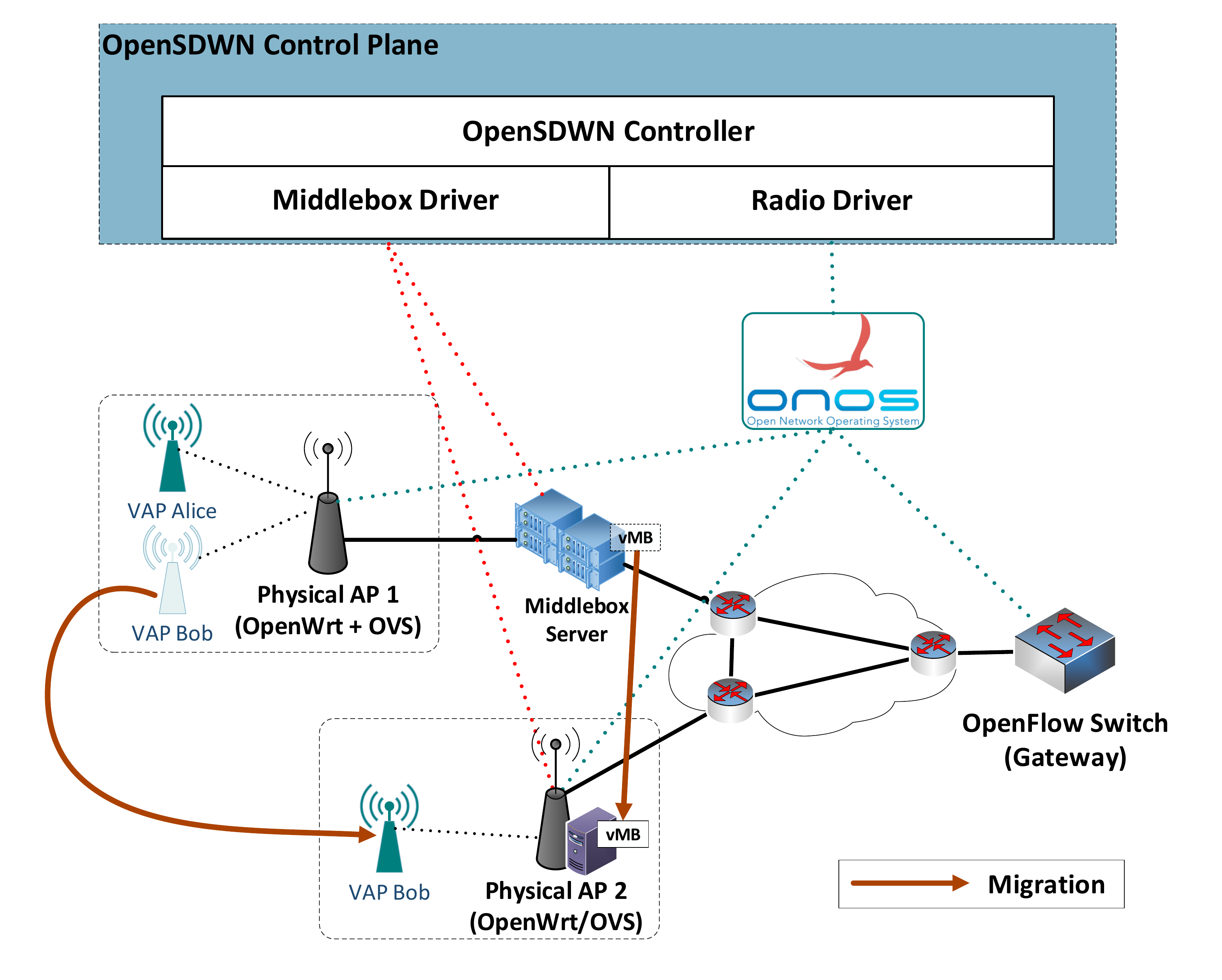}
	\end{adjustbox}
	\caption{NFV/SDN architecture for WiFi networks.}
	\label{fig:lvaps}
\end{figure}

Regarding a WiFi network, the studies related to WLANs leverage the Virtual Access Point (VAP) abstraction by moving the MAC layer or middleboxes processing to the Cloud. When associated with the wireless network, each client acquires a VAP that will be dedicated, independent of the client migrating from one access point to another (handover), as shown in Figure \ref{fig:lvaps}. In this example, the Physical AP 1 allocates two VAPs to two clients, Bob and Alice. If Bob moves toward another AP, his VAP will also be migrated and deployed to the new AP. 

Shulz-Zander et al. (2015) \cite{schulz-zander_opensdwn:_2015,8016637} proposed the OpenSDWN, an NFV/SDN approach to implement per client access points and virtual middleboxes. Figure \ref{fig:lvaps} shows our detailed view of this architecture. For the access point case, the authors created an extension to Odin \cite{Suresh:2012:TPE:2342441.2342465}, called Light Virtual Access Point (LVAP). An LVAP uses SDN applications to abstract some functionality of the 802.11 Access Point, such as authentication, handoff, and client associations. A physical AP supports multiple LVAPs, one for each client (which receives a unique BSSID). Therefore, a certain LVAP serves as a dedicated link between its client and infrastructure. The authors also implemented virtual middleboxes (e.g., firewall), which can be deployed either on a Middlebox Server or at the access point itself and can integrate them with LVAPs using virtual networks. A service differentiation mechanism (DPI-based) tries to identify and classify flows to redirect traffic to the correct vMiddleboxes (vMB). OpenSDWN Controller performs all the above functionalities, using the Floodlight \cite{schulz-zander_opensdwn:_2015} and ONOS \cite{8016637} as SDN controllers. Such an abstraction allows the seamless mobility with the migration of both LVAPs and vMBs among APs.  

In \cite{vestin_qos_2015}, the authors extended CloudMAC framework \cite{6477567} to provide QoS for VAPs. In the proposed solution, the VAP is responsible for the MAC layer management frames (e.g., beacons, probes request/response) and runs as a VM in a Cloud environment. The physical AP redirects these frames to the destination VAP, using OpenFlow rules. The QoS mechanism implements VAP traffic prioritization using different queue management strategies (e.g., Stochastic Fair Queueing) on all Open vSwitches between the APs and VAPs. The OpenDaylight Controller manages both traffic redirection and prioritization. Seamless handovers can be achieved by just changing the SDN forwarding rules. 

\subsubsection{\textbf{Wireless Mesh Networks (WMN)}}

The only study found for WMN proposes the Urban-X, an NFV/SDN solution for dense urban scenarios \cite{kim_toward_2015}. In this case, the authors create a multi-radio cognitive mechanism to dynamically self-adapt when there are variations in the interference conditions on the WiFi channels. For this, each mesh node includes an OF Agent component that supports OpenFlow and allows the instantiation of VNFs. The VIM component controls all OF Agents. It uses an OpenFlow Controller to establish a path for end-to-end connectivity, including one or more mesh nodes. For this, VIM uses the OF Agents to monitor link status to configure the path with minimum latency. At the same time, the VIM instantiates TCP accelerators as VNFs at each mesh node included in the path. Thus, the Urban-X improves TCP throughput to the mesh clients.        

\subsection{Wireless and Mobile Networks}

A new generation of mobile network technologies appears every ten years. The first generation (1G) came in the mid-1980s with analog cellular networks. The second generation (2G) began in the mid-1990s and started the era of digital mobile phones encompassing technologies such as CDMA, TDMA, GSM, GPRS, and EDGE. The third-generation (3G) emerged in the late 1990s introducing the use of packet switching rather than circuit switching for data transmission. 3G technologies such as the Universal Mobile Telecommunications System (UMTS) achieves high connection speeds (up to 42 Mbit/s downlink), making it possible to use multimedia applications. The fourth generation (4G) with its Long Term Evolution (LTE) appeared in the mid-2010s with the aim of providing speed improvements up to 10-fold over the existing 3G technologies.

It is worth emphasizing that every new generation tries to address service and network requirements not met by its predecessors. One of the current challenges for mobile networks is how to handle the ever-increasing traffic volume. According to Cisco VNI \cite{cisco}, mobile traffic volume grew 63\% in 2016, reaching an average of 7.2 exabytes per month. Such research forecasts an increase in this monthly traffic volume by seven times in the future, reaching the mark of 49 exabytes. To address this growth, mobile operators are investing in infrastructure, thus increasing OPEX/CAPEX costs and management complexity. In this context, 5G networks aim at addressing the following demands \cite{7169508}: improved data rate, decreased latency, and increased capacity for consistent QoS/QoE. 

Currently, the standardization of 5G networks is still a work-in-progress \cite{itu5g}. The International Telecommunications Union (ITU\footnote{ITU website: http://www.itu.int}) will be the specialized agency responsible for publishing the final standard in mid-2020's, which is also referenced as International Mobile Telecommunications (IMT)-2020. The 3rd Generation Partnership Project (3GPP\footnote{3GPP website: www.3gpp.org/}) is the standard body that unites several mobile industries with the objective of elaborating and submitting a proposed specification to the ITU (mid-2018's) to be part of the IMT-2020 standard.

Both industrial and academic researchers have also put research efforts on the architectural components of 5G networks. For instance, Verizon created a 5G Tech Forum (5GTF) in September 2015 \cite{5gtf}. The 5GTF is a vital initiative where major vendors such as Verizon, Cisco, Ericsson, Nokia, and Apple work together to develop early 5G specifications and then contribute to the 3GPP. 5GTF published its first specification release in July 2016.  

Also, the European Union funded 5G Public-Private Partnership (5GPPP). 5GPPP is a joint initiative between the European Commission (EC) and the European ICT industry to develop solutions, architectures, and standards to put Europe in the leadership position for the 5G networks \cite{5gppp}. The 5GPPP has been supporting different projects in its first\footnote{5G PPP Phase I Projects - https://5g-ppp.eu/5g-ppp-phase-1-projects/} and second\footnote{5G PPP Phase II Projects - https://5g-ppp.eu/5g-ppp-phase-2-projects/} phases such as METIS and SELFNET. Those projects focus on research topics ranging from physical infrastructure to overall architecture, virtualization, network management, and software networks. As a result, 5GPPP has published several specifications, including a view on the 5G architecture \cite{5gppparchitecture}.  

A 5G infrastructure must provide features that support different types of vertical business such as Automotive (e.g., car manufacturers), eHealth (e.g., health industry), Energy (e.g., power companies), Factories (e.g., IoT technology providers), Media \& Entertainment (e.g., content providers) \cite{5gppparchitecture}. All of those markets encompass different types of use cases (e.g., automated driving, robotics for remote surgery, on-site live event experience, etc.) that have their characteristics (e.g., data traffic patterns, mobility support, etc.) and requirements (e.g., throughput, latency, etc.). An exhaustive list of case studies for 5G can be found in \cite{5gpppcs5g}. Furthermore, some performance requirements have been enumerated for such new generation of mobile networks \cite{7169508,6815890}:  

\begin{itemize}
\item 10 to 100 times higher data rate (1 to 10 Gbps);
\item 5 times reduced end-to-end latency (less than 5 milliseconds);
\item 1000 times higher capacity (9 Gigabytes per hour in busy period and 500 Gigabytes per month per subscriber);
\item 10 to 100 times higher massive number of connections (300,000 connections per access point);
\item 10 times extended battery life;
\item Availability of 99.999\%.
\end{itemize}

\begin{figure}[htbp]
	\centering
	\begin{adjustbox}{width=0.7\textwidth}
		\includegraphics[]{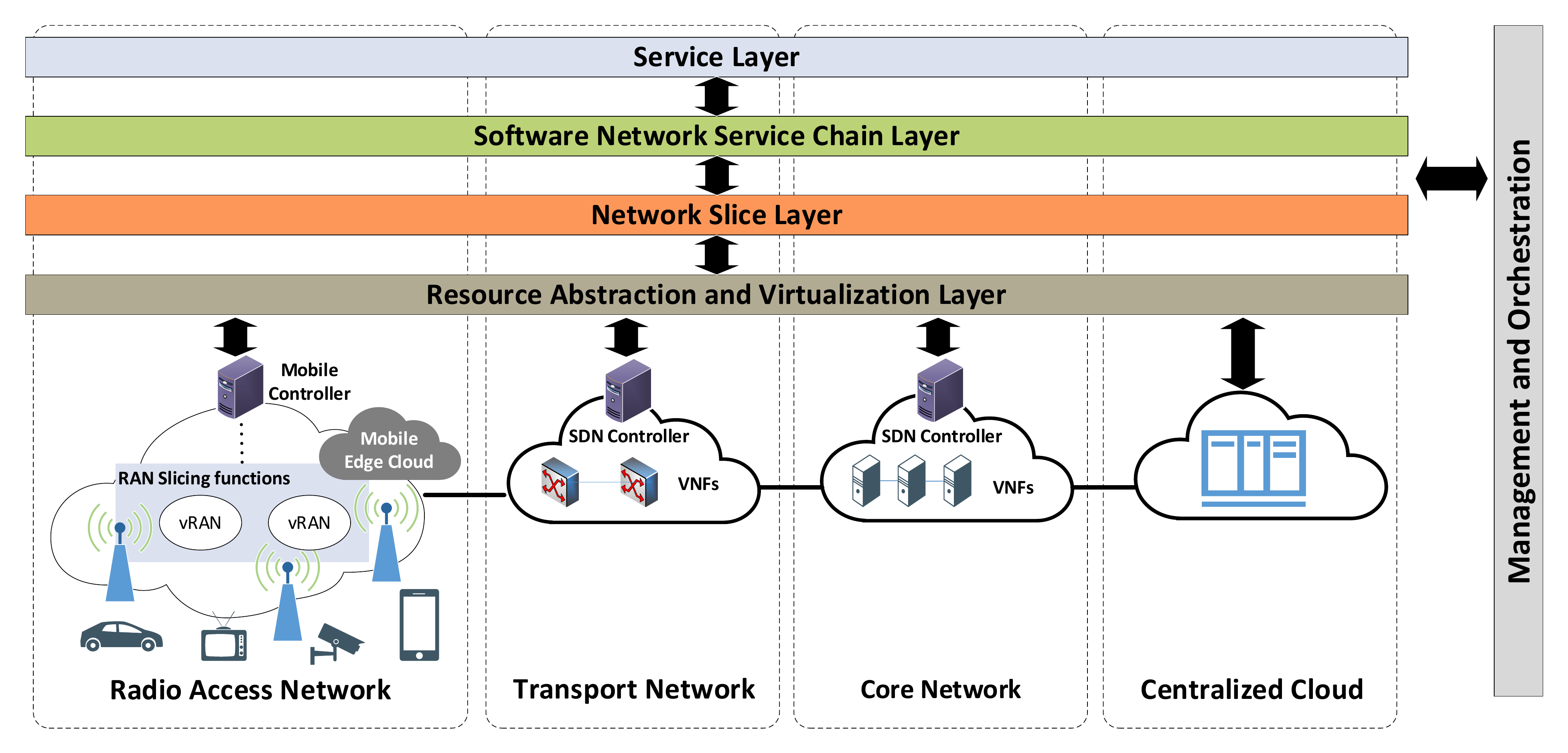}
	\end{adjustbox}
	\caption{Software network technologies for 5G infrastructures \cite{5gppparchitecture}.}
	\label{fig:soft5g}
\end{figure}

To support such heterogeneity in the use cases as well as to meet the performance requirements, new 5G technologies will impact the entire mobile network including mobile devices; radio access, transport, and core networks; and the cloud (local, regional, or global). Figure \ref{fig:soft5g} shows this type of infrastructure. A 5G architecture should enable speed, agility, and cost-efficiency when delivering new services such as those in the context of the Internet of Things (IoT) and Smart Cities. 5G networks should also provide multi-tenancy, multi-service, and multi-domain support. To this end, the infrastructure providers must allocate logical networks (accessible by northbound APIs), so-called network slices (Network Slice layer), or for mobile operators or service providers, who in turn can create their own slices or services (Software Network Service Chain and Service layers). To build logical networks, the infrastructure providers will have to deploy end-to-end resource, infrastructure (Resource Abstraction and Virtualization layer), and service orchestration functions to reserve appropriate computing and network resources from different administrative domains, keeping QoS tailored to user demand \cite{5gppparchitecture}.  

Finally, in the Radio Access Network (RAN), 5G architecture should operate in a broad spectrum range with a diverse variety of characteristics, provide efficient transmission and data processing, support the coexistence of different radio access technologies (5G, LTE, and Wi-Fi) and be energy efficient. In this case, techniques such as Mobile Edge Computing (MEC) can be used as it allows pushing the services to the RAN with the objective of meeting the ultra-low latency and higher-speed requirements \cite{5gppparchitecture,huawei5g}.

The key enablers to achieve these functions are virtualization, softwarization, and programmability features, which can deliver the suitable level of flexibility in 5G networks. The use of NFV and SDN technologies will play a significant role in 5G networks, since they allow the network programmability and the fast delivery of new services, enabling network slicing and MEC implementation and orchestration \cite{5gpppsofnet5g}. In this context, several NFV/SDN architectures have been designed to overcome most of these challenges. We highlight some prospective solutions in the following sub-sections.

\subsubsection{\textbf{Mobile Network Function Virtualization}}

These architectures use cloud computing to assist mobile network virtualization. The goal is to provide a virtualization and communication platform for mobile network services as a way to deliver a flexible and scalable environment.

Regarding 3G services, \cite{kyung_software_2015} proposed the Software Defined Transitional Networking (SDTN), an NFV/SDN architecture to support legacy service integration in 4G networks. The authors assumed LTE as the underlying network (SDTN data plane). 3G functions are VNF instances that replace the following components: Serving General Packet Radio Service (GPRS) Support Node (SGSN), Gateway GPRS Support Node (GGSN), and Home Location Register (HLR). The Edge Controller acts as an SDN Controller mapping the actions performed by the virtualized 3G functions to the physical 4G network (using the 4G forwarding control plane). The Edge Controller coordinates the redirection of network services traffic to VNFs by programming edge switches (using OpenFlow).

When considering 4G services, most of the studies that focus on the mobile core network include the virtualization of Evolved Packet Core (EPC), such as: Serving Gateway (SGW) \cite{poc_etsi_26,poc_etsi_27,haleplidis_forces_2014,basta_sdn_2014,nguyen_slicing_2014,costa-requena_sdn_2015,medhat_multi-tenancy_2015,ahmad_new_2016,7997391}, Packet Data Network Gateway (PGW) \cite{poc_etsi_26,poc_etsi_27,haleplidis_forces_2014,an_virtualization_2014,basta_sdn_2014,nguyen_slicing_2014,costa-requena_sdn_2015,medhat_multi-tenancy_2015,ahmad_new_2016,7997391}, Mobility Management Entity (MME) \cite{poc_etsi_26,poc_etsi_27,nguyen_slicing_2014, costa-requena_sdn_2015,medhat_multi-tenancy_2015,ahmad_new_2016}, Home Subscriber Server (HSS) \cite{poc_etsi_26,costa-requena_sdn_2015,ahmad_new_2016}, and Policy and Charging Rules Function (PCRF) \cite{costa-requena_sdn_2015,ahmad_new_2016}.

In \cite{poc_etsi_26} and \cite{costa-requena_sdn_2015}, the authors proposed a PoC to evaluate the virtualization of EPC components (vSGW, vPGW, and vMME) as VNFs over an NFV/SDN architecture. The testbed comprised eNodeBs (emulator or eNodeB model Flexi Zone from Nokia Networks) interconnected with a Cloud data center through OpenFlow switches with MPLS support (Coriant Oy 8615 Smart Router). All EPC VNFs run in a Cloud environment and were implemented using the following tools: eMME SW module (Aalto University) for vMME, open source nwEPC\footnote{https://www.openhub.net/p/nwepc} for vSGW and vPGw, and an SQL database fo vHSS. A Ryu Controller coordinates the OpenFlow switches providing NFVI connectivity, eNodeB and VNFs interconnection and QoS support using MPLS tagging.

In addition, several studies proposed NFV/SDN architectures for virtualization of SGi-LAN services \cite{poc_etsi_13,poc_etsi_23,poc_etsi_34,gronsund_solution_2015}. Serving Gateway interface (SGi) interconnects mobile packet core and external IP networks. In a 4G network, SGi runs between PGW and a Packet Data Network (PDN) being responsible for ensuring the intercommunication performance and reliability. For this, SGi encompasses some services, such as Deep Packet Inspection (DPI), firewall, NAT, TCP optimization, and several caches. Gronsund et al. (2015) \cite{gronsund_solution_2015} proposed to replace the SGi elements for a physical OpenFlow Switch. The VNFs (TCP and Video Optimizer, firewall, HTTP content filter, etc.) run in a Cloud data center environment with RHEL OpenStack (Red Hat). The OpenStack coordinates the OpenDaylight Helium Controller to create OpenFlow rules in the switch to redirect and load balancing traffic to the VNFs. To keep track of VNF instances (in constant quantity variation for elasticity purposes) the OpenStack uses the LISP mapping service of OpenDaylight. 

\subsubsection{\textbf{Network Slicing}}

\begin{figure}[htbp]
	\centering
	\begin{adjustbox}{width=0.5\textwidth}
		\includegraphics[]{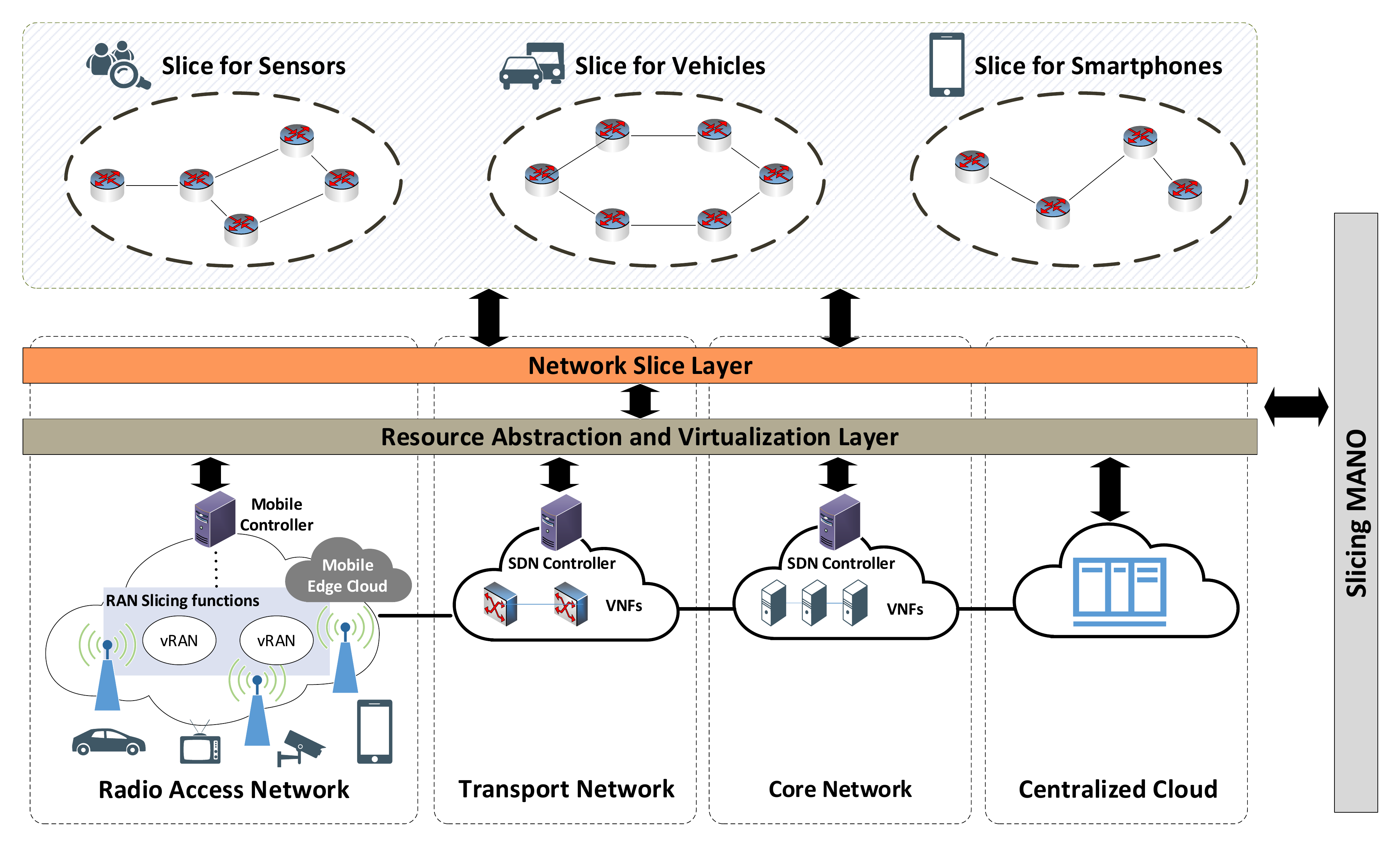}
	\end{adjustbox}
	\caption{Conceptual illustration of network slicing.}
	\label{fig:netslice}
\end{figure}

Network Slicing refers to the partitioning of a certain physical infrastructure, composed of both network and computational resources, into multiple logical networks, called network slices \cite{7926921}. Figure \ref{fig:netslice} shows that each slice is a self-contained network with its own virtual resources created on top of the underlying infrastructure. It can be designed and optimized for a particular mobile operator or service provider. 

When compared to traditional physical networks, Network Slicing have the following advantages \cite{7926921}: i) customization of logical networks according to service requirements; ii) on-demand provisioning to scale resources up or down as conditions change, and iii) network resource isolation for improved security and reliability. 

Network Slicing aims at providing efficient resource sharing, traffic differentiation per slice, and management and protection tools \cite{5gppparchitecture}. NFV and SDN technologies are capable of providing the flexibility required in this context \cite{5gpppsofnet5g}.  

A use case for the use of NFV/SDN architectures in Network Slicing is for the creation of SDN-enabled Virtual Tenant Networks (VTNs). VTNs are virtual networks deployed to different tenants in an isolated way (independent of underlying physical network resources) to support specific Quality of Service (QoS) and Service Level Agreement (SLA) requirements. There is a trend to use SDN in the creation of virtual networks. By enabling network programmability, SDN renders the abstraction necessary for its use as a network hypervisor. In the case of SDN-Enabled VTNs, one or many SDN controllers create a VTN (called an Infrastructure SDN Controller), while a new SDN controller is instantiated to manage this VTN (called a Tenant SDN Controller).   

When an SDN-Enabled VTN deployment takes place, the respective Tenant SDN Controller is manually installed and configured on a dedicated server, which can be a long process. NFV/SDN architectures can be used to virtualize tenant SDN Controllers and provide fast and dynamic VTN provisioning.  

The works of Munoz and Vilalta \cite{munoz_integrated_2015,munoz_sdn/nfv_2015,vilalta_sdn/nfv_2015,vilalta_multi-tenant_2015,vilalta_multitenant_2016,7767521} proposed an NFV/SDN solution for fast and dynamic deployment of SDN-Enabled Virtual Tenant Networks over multiple Data Centers and WAN domains. Their solution aims at providing geographically distributed cloud services with specific QoS and SLAs.  

\begin{figure}[htbp]
	\centering
	\begin{adjustbox}{width=0.5\textwidth}
		\includegraphics[]{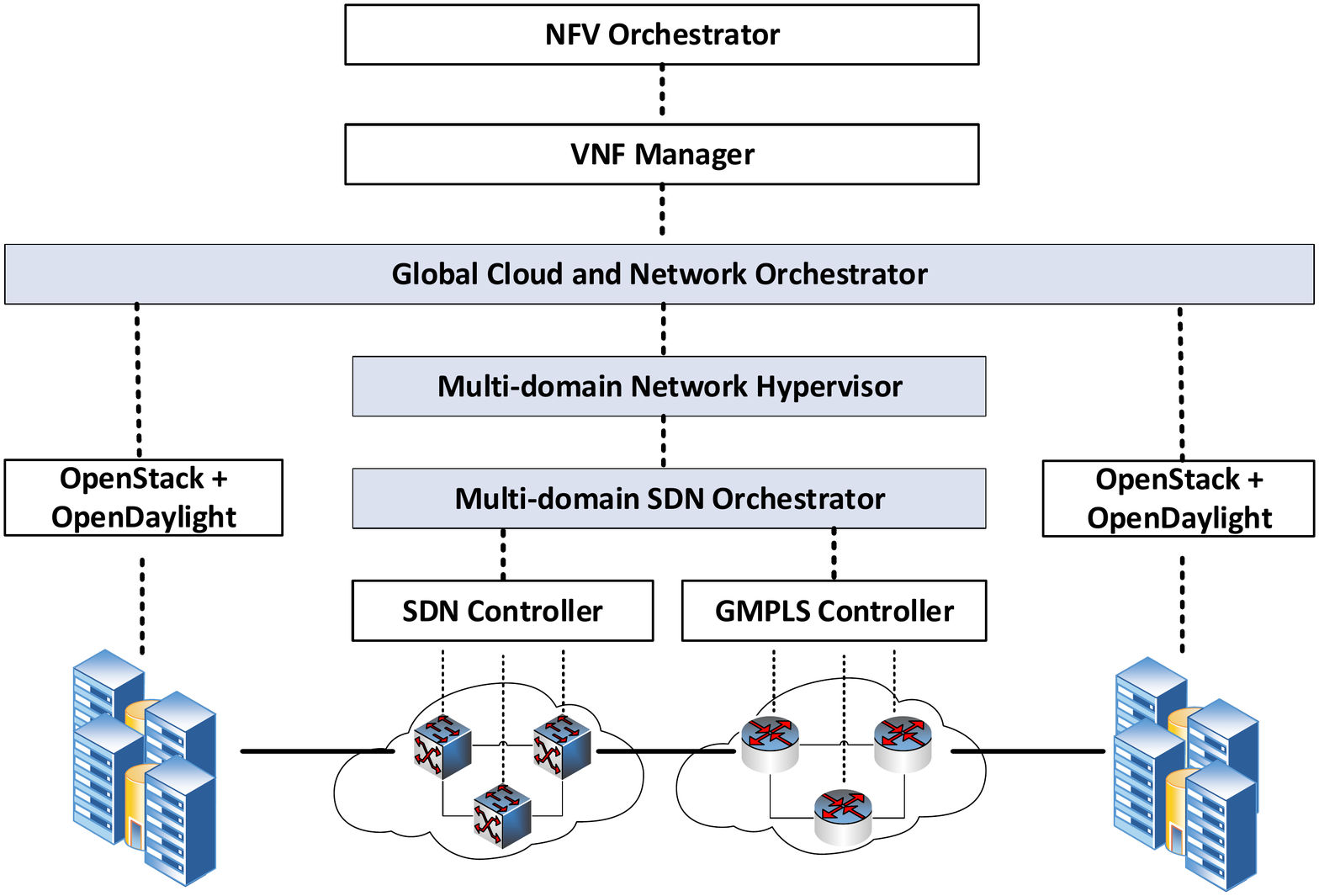}
	\end{adjustbox}
	\caption{An NFV/SDN architecture design for SDN-enabled VTNs \cite{munoz_integrated_2015}.}
	\label{fig:munoz}
\end{figure}

In \cite{munoz_integrated_2015}, the authors used  NFV and Cloud to virtualize tenant SDN Controllers (OpenDaylight or Floodlight) to control the underlying SDN-enabled VTNs and provide fast and dynamic VTN provisioning (see Figure \ref{fig:munoz}). They used OpenStack as VIM for each Data Center and an OpenDaylight controller to interconnect a virtual tenant SDN Controller with its respective VTN. 

To create the VTNs, they used the Multidomain SDN Orchestrator (MSO) mechanism as a Network Operating System (NOS). The MSO creates an abstraction over multiple domains including different transport network technologies thus enabling the composition of end-to-end services over heterogeneous WAN networks. Also, the authors use the Multidomain Network Hypervisor (MNH) to create end-to-end SDN-enabled VTNs, over the abstraction provided by MSO. Using the Global Cloud and Network Orchestrator, a VIM mechanism, this architecture integrates geographically distributed Data Centers and multiple WAN domains, providing a unified cloud and network operating system for the creation of end-to-end NFV services over VTNs.  

Several research papers focus on providing Network Slicing \cite{akyildiz_wireless_2015,mwangama_towards_2015,nguyen_slicing_2014,medhat_multi-tenancy_2015,7926921} for the new generation of mobile network. The 3GPP has identified Network Slicing as one of the key technologies to achieve the goals in 5G Networks \cite{7926921} since it is a potential solution to enable suitable flexibility to address the specific requirements of different use cases. In this scenario, mobile operators could share the same physical network substrate, adding their virtual networks with their services (e.g., 3G, 4G services) and a centralized management plane, creating the so-called Mobile Virtual Network Operators (MVNO). These logical networks must be isolated from each other as a way to maintain privacy between operators. In this case, NFV provides the mobile network services per operator as VNFs and, in turn, SDN creates the slice as well as establishes network functions interconnectivity.  

Mwangama et al. (2015) \cite{mwangama_towards_2015} designed an NFV/SDN architecture to support MVNOs in a federated cloud environment. A prototype was implemented using the non-open source FOKUS OpenSDNCore Orchestrator \cite{opensdncore} to coordinate the network services between MVNOs. The Orchestrator uses OpenStack as VIM to create the virtual tenant networks and to instantiate the following VNFs per mobile operator: EPC (non-open source FOKUS OpenEPC\footnote{http://www.openepc.com/} platform), IMS – IP Multimedia Subsystem (open source FOKUS OpenIMSCore\footnote{http://www.openimscore.org/}), M2M (non-open source FOKUS OpenMTC\footnote{http://www.open-mtc.org/}).  

Li et al. (2017) \cite{7926921} proposed a three-layer Network Slicing framework model for 5G networks considering NFV and SDN technologies. The bottom layer is the 5G Software-Defined Infrastructure (5G-SDI) which comprises multiple administrative and physical domains (e.g., RAN, transport and core networks, etc.) with SDN-based control and management. Their SDN-based approach uses hierarchically organized SDN controllers to provide abstraction and distributed dynamic allocation of resources. Furthermore, RAN and MEC can be deployed to enable a cloud-based infrastructure. The Virtual Resource layer creates network slices with virtual resources (radio, computing, and network) and VNFs that are customized to meet the requirements of different types of services. The Application and Service layer includes the per-tenant services (e.g., connected vehicles, virtual reality, etc.) that will use these slices to perform their functionalities. Also, the life cycle of network slices is managed and orchestrated by the Slicing MANO that acts as VIM, VNF Manager, and Slice Orchestrator.   

Recently Munoz and Vilalta (2016 and 2017) \cite{7494060,7718525,7767629,7858117,7980775,7561003} have adapted the previously defined NFV/SDN architecture \cite{munoz_integrated_2015} to 5G scenarios, including the entire mobile network (e.g., radio access network) for fast and dynamic deployment of MVNOs. Besides, in \cite{7980775,7561003} the authors proposed the ADRENALINE Testbed for 5G and IoT services on top of an NFV/SDN platform.

\subsubsection{\textbf{Mobile Edge Computing (MEC)}} \label{sub:mec}

\begin{figure}[htbp]
	\centering
	\begin{adjustbox}{width=0.7\textwidth}
		\includegraphics[]{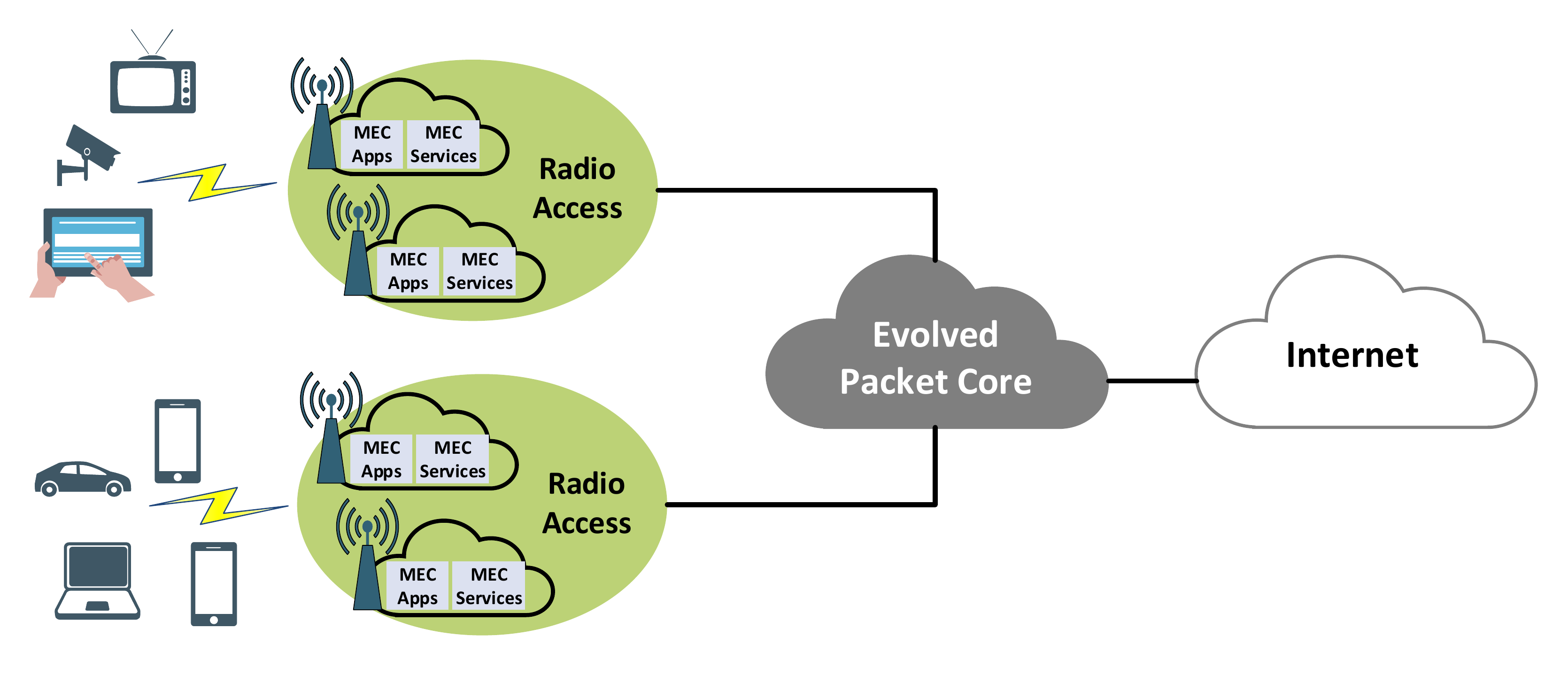}
	\end{adjustbox}
	\caption{Mobile Edge Computing scenario.}
	\label{fig:mec}
\end{figure}

Mobile Edge Computing (MEC) or Multi-access Edge Computing has been a trend in mobile networks. Like NFV, the MEC architecture has been standardized by ETSI through Group Specification (GS) MEC \cite{etsi_mec_002,etsi_mec_003} since 2016. MEC provides IT and Cloud Computing capabilities within the Radio Access Network (RAN). For this, a set of computer and storage resources (e.g., data centers, clusters, etc.) are deployed at the edges of a mobile operator's network to assist the core data center in supporting computing and communication (see Figure \ref{fig:mec}) \cite{DBLP:journals/corr/RomanLM16}. MEC focuses on delivering the services closest to the user, as a way to meet certain critical application (e.g., video analytics, Internet-of-Things, augmented reality, and data caching) requirements that are not supported only by Cloud Computing, such as high bandwidth, low latency and jitter, context awareness, and mobility support.

According to 5G-PPP, MEC is vital technological component to enable 5G networks \cite{5gpppvision5g}. NFV/SDN architectures are in line with current trends for MEC solutions. Because it is a new technology, only a few studies have been found in this SLR. As an example, the EU H2020 SELFNET project \cite{selfnetproject} proposes the design and implementation of an Autonomic Management Framework for 5G networks, using technologies such as SDN, NFV, Self-Organizing Network (SON), Cloud Computing, and Artificial Intelligence. This framework aims at reducing OPEX and at improving QoE of the end users, addressing the following self-organizing capabilities: i) self-protection against distributed cyber-attacks, ii) self-healing against network failures, and iii) self-optimization of the network traffic. In this context, Neves et al. proposed a SELFNET approach to support SFC in MEC scenarios \cite{neves_selfnet_2016,NEVES2017229}, to meet 5G requirements defined by the 5G-PPP initiative \cite{5gppp}. They considered a federated cloud infrastructure (i.e., multiples edge NFVI-PoP and a core NFVI-PoP) to provide IT and network resources to execute VNFs that support some management elements and network services. The WAN Infrastructure Management (WIN) uses SDN Controllers to provide connectivity between edge NFVI-PoPs and the core NFVI-PoP through the creation of virtual tenant networks.

\section{TAXONOMY OF NFV/SDN ARCHITECTURES DESIGN}
\label{sec:taxonomy}

This section provides support to answer the third research question, by describing a taxonomy to organize the various decision-making levels for the design of NFV/SDN architectures.

\begin{figure}[htbp]
\centering
\begin{adjustbox}{width=1\textwidth}
\includegraphics[]{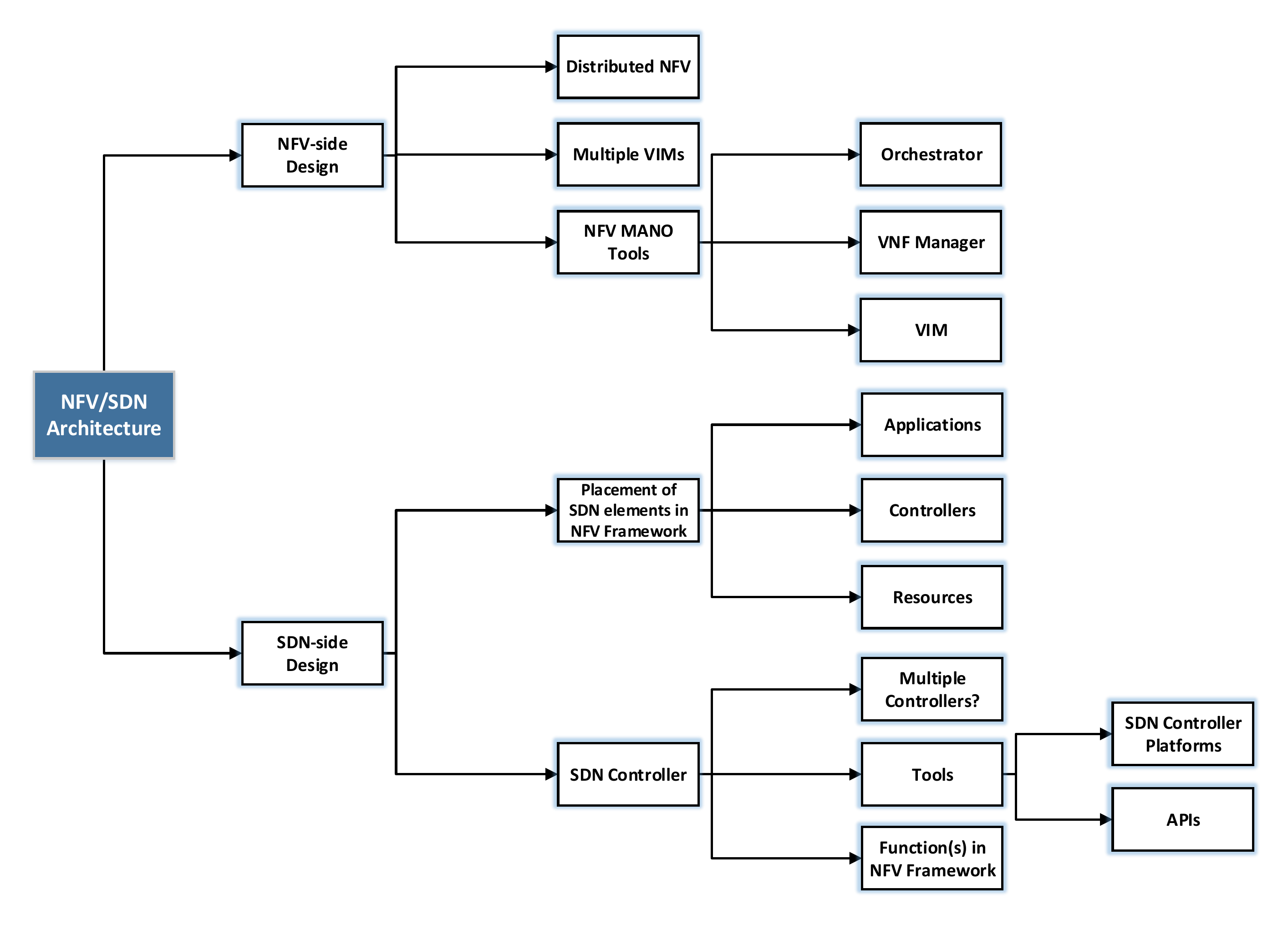}
\end{adjustbox}
\caption{A taxonomy to classify NFV/SDN architectures design.}
\label{fig:taxonomy}
\end{figure}

Figure \ref{fig:taxonomy} depicts our proposed taxonomy that provides a conventional architectural design for using SDN in an NFV framework. It was derived from architectures and implementations found in the selected studies and published NFV/SDN reference architectures \cite{etsi_sdn_in_nfv,verizon}. This taxonomy is useful as a guide for simplifying the work of researchers when studying NFV/SDN architectures or providing new solutions. 

In this taxonomy, the NFV/SDN architectures design was divided into two sides: NFV-side and SDN-side. In the NFV-side, we must decide whether or not to use two features inherent to the architecture design. We describe these features as follows.

\begin{description}
\item[Distributed NFV (D-NFV):] In D-NFV, the MANO framework places Virtual Network Functions (VNFs) where they could be most efficiently and economically be deployed, such as in data centers, forwarding devices, or the CPEs \cite{Gittik2014}.
\item[Multiple VIMs:] A designer could place Multiple VIMs in different NFVI-PoPs to support the multi-domain administration or in the same NFVI-PoP to provide scalability and performance \cite{etsi_sdn_in_nfv}.  
\end{description}

As far as we are concerned with NFV, it is important to identify the NFV Management and Orchestration (MANO) tools. Tools such as OpenMANO \cite{openmano} and OpenBaton \cite{openbaton} can provide complete solutions for MANO. On the other hand, the OpenStack enables VIM implementation to provide support for existing or new VNF Managers and NFV Orchestrators.

On the SDN-side, a first step is the placement of SDN elements in the NFV Framework \cite{etsi_sdn_in_nfv}. These elements are described below:

\begin{description}
\item[SDN Resources:] Comprise of both physical and virtual switches and routers;
\item[SDN Controllers:] Responsible for controlling the SDN resources, determining the behavior of network traffic; 
\item[SDN Applications:] Interfaces with one or multiple SDN controllers to enforce high-level network policy, such as firewall, network address translation, QoS, and network management.
\end{description}

We list some possible locations for the placement of SDN Resources in the NFV Framework, as follows  \cite{etsi_sdn_in_nfv}:

\begin{itemize}
\item Physical switch or router;
\item Virtual switch or router;
\item E-switch, software-based SDN-enabled switch in a server NIC;
\item Switch or router as a VNF.
\end{itemize}

There are also some possible locations for the placement of SDN Controllers in NFV Framework \cite{etsi_sdn_in_nfv}. They are the following:

\begin{itemize}
\item Merged with the Virtualized Infrastructure Manager (VIM);
\item Virtualized as a VNF;
\item As part of the NFVI and not as a VNF;
\item As part of the OSS/BSS;
\item As a Physical Network Function (PNF).
\end{itemize}

Finally, some locations for the placement of SDN Applications in NFV Framework are listed below \cite{etsi_sdn_in_nfv}:

\begin{itemize}
\item As part of a PNF;
\item As part of the VIM;
\item Virtualized as a VNF;
\item As part of an EMS;
\item As part of the OSS/BSS.
\end{itemize}

There are also some architectural decisions to be made when one considers the SDN Controller, as follows:

\begin{itemize}
\item \textbf{Does the solution implement multiple SDN Controllers? If so, what is the main objective?} Multiple SDN Controllers are hierarchically distributed to provide performance, scalability, reliability, administrative domains interaction, or Network as a Service (NaaS) management in an NFV Framework.
\item \textbf{What is the function of SDN Controller in the NFV Framework?} Network Connectivity in the NFVI, Control of Virtual Networks, Interconnecting VNFCs, and Interconnecting VNFs are some of these functions (extracted from the studies selected in this SLR).
\end{itemize}

Finally, we must identify the SDN Controller tools, including its underlying software and the used bound interfaces (at South, North, West, and East). As SDN Controllers we can cite: OpenDaylight \cite{opendaylight}, Floodlight \cite{floodlight}, ONOS \cite{onos}, Ryu \cite{ryu}, and POX \cite{pox}.

The next Sections (\ref{sec:nfvside} and \ref{sec:sdnside}) aim to answer research question 3, using this taxonomy to organize the description and the differences between the NFV/SDN solutions extracted from articles selected in SLR. 
\section{NFV-SIDE DESIGN}
\label{sec:nfvside}

This section uses the NFV-side of the taxonomy described in Section \ref{sec:taxonomy} to organize the NFV/SDN solutions extracted from articles selected in the SLR.

We organized the studies according to how they implemented the components of MANO (i.e., the NFV Orchestrator, the VNF Manager, and the Virtualized Infrastructure Manager). For each component, the research studies are classified as follows:

\begin{itemize}
\item Real Implementation: The study proposes and implements its own component; 
\item Theoretical: The study proposes its own component, but it is not implemented;
\item Vendor-specific: The study uses a proprietary tool to implement the component.
\end{itemize}

\begin{figure}[htbp]
	\centering
	\begin{adjustbox}{width=\textwidth}
		\includegraphics[]{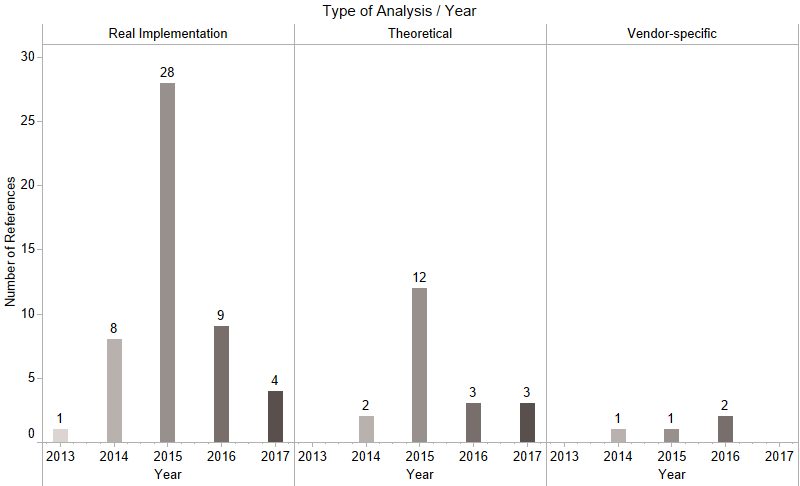}
	\end{adjustbox}
	\caption{Number of studies per Type of Analysis/Year.}
	\label{fig:prop}
\end{figure}

The majority of the articles were classified as \textit{Real Implementation} (see Figure \ref{fig:prop}). These studies adopt modern tools to assist in the implementation of solutions, mainly the VIM component (e,g., OpenStack \cite{openstack}). However, it is worth mentioning that most of them implement the orchestration functions without the support of existing NFV MANO frameworks, such as OpenStack Tacker, OpenMANO, OpenBaton, Open-O, ECOMP, Hurtle, and the like. On the other hand, three papers \cite{medhat_multi-tenancy_2015,carella_cross-layer_2015,mwangama_towards_2015} developed their solutions on top of the OpenSDNCore Orchestrator \cite{opensdncore}, from Fraunhofer FOKUS Institute. Last, we highlight that the Vendor-specific solutions are only used in some PoCs from ETSI NFV ISG.

Furthermore, some studies have provided architectures that deal with Distributed NFV (D-NFV) and Multiple VIMs. With D-NFV we could place VNFs wherever they may be most effective (performance and scalability) and least expensive. On the other hand, Multiple VIMs are often used to perform management of several administrative domains or NFVI-PoPs. In this scenario, VIMs are hierarchically distributed. So-called secondary VIMs are responsible for managing NFVI-PoPs. The primary VIM controls the secondary VIMs to create an abstraction layer on all NFVI-PoPs and performs a centralized management. Such a hierarchy enables the creation of end-to-end network services, involving multiple domains (e.g., Cloud and WAN).

\begin{table}[ht]
	\caption{List of studies addressing Distributed NFV and/or Multiple VIMs.}
	\label{tab:q31}
	\begin{minipage}{\columnwidth}
		\begin{center}
			\begin{tabular}{p{3cm}p{10cm}}
				\toprule
				\textbf{Implementation} & \textbf{Studies} \\ \midrule
				Distributed NFV & \cite{poc_etsi_34,shen_vconductor:_2015,cerrato_toward_2015,soares_toward_2015, soares_cloud4nfv:_2014,vilalta_sdn/nfv_2015, vilalta_transport_2015,vilalta_multitenant_2016,mohammadkhan_protocols_2015,munoz_sdn/nfv_2015,munoz_integrated_2015,vilalta_multi-tenant_2015,carella_cross-layer_2015,sonkoly_unifying_2015,neves_selfnet_2016,lombardo_open_2015,basta_sdn_2014,kyung_software_2015,mamatas_service-aware_2015,rossem_deploying_2015,7926921,7945849,7767521,7494060,7718525,7767629,7858117,7980775,7561003,7579021,NEVES2017229} \\
				Multiple VIMs & \cite{poc_etsi_16,poc_etsi_34,basta_sdn_2014,munoz_integrated_2015,Szabo2015,neves_selfnet_2016,cerrato_toward_2015,soares_toward_2015,soares_cloud4nfv:_2014,mohammadkhan_protocols_2015,shen_vconductor:_2015,batalle_implementation_2013,carella_cross-layer_2015,kyung_software_2015,vilalta_sdn/nfv_2015,vilalta_transport_2015,vilalta_multi-tenant_2015,vilalta_multitenant_2016,munoz_sdn/nfv_2015,sonkoly_unifying_2015,rossem_deploying_2015,7926921,7945849,7767521,7494060,7718525,7767629,7858117,7980775,7561003,7579021,NEVES2017229} \\
				\bottomrule
			\end{tabular}
		\end{center}
	\end{minipage}
\end{table}		

\begin{figure}[htbp]
	\centering
	\begin{adjustbox}{width=\textwidth}
		\includegraphics[]{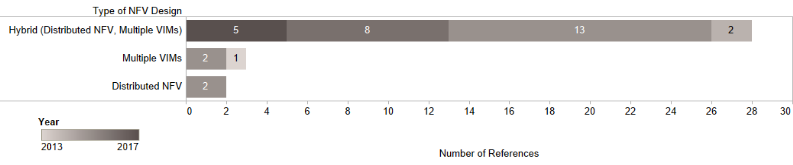}
	\end{adjustbox}
	\caption{Number of studies addressing Distributed NFV and/or Multiple VIMs.}
	\label{fig:nfv-des}
\end{figure}

Table \ref{tab:q31} lists the studies that implement Distributed NFV or Multiple VIMs. It is worth mentioning that most of the studies implemented both designs (see Figure \ref{fig:nfv-des}). As an example, \cite{shen_vconductor:_2015} proposed the vConductor, a Cloud CPE (see Section \ref{sub:vcpe}) solution for automation of multi-tenant virtual network provisioning. vConductor deploys all enterprise network functions as VNFs in a Cloud domain comprised of multiple data centers. By using a User Portal, the customers can acquire new network functions and define how their VNFs must be chained. A virtual tenant network (VTN) is established connecting the enterprise CPE and the Cloud infrastructure through an OpenFlow-enabled WAN domain. Each data center in a Cloud domain uses an OpenStack as a management platform (secondary VIM). Further, an OpenDaylight Controller (secondary VIM) updates the OpenFlow rules required for VTN management in WAN domain. vConductor acts as NFVO, VNFM, and primary VIM, controlling the multiple secondary VIMs. Finally, vConductor includes the Virtual Network Life Cycle Manager (VNLM) to creates a D-NFV scenario. VNLM implements a multi-objective resource scheduling algorithm (MORSA) that uses a genetic algorithm to provides near-optimal placement of VNFs over different data centers.

However, there are also works that implement only one of these scenarios. As an example of D-NFV scenarios without multiple VIMs, \cite{lombardo_open_2015} proposed an NFV/SDN architecture, called NetFATE (Network Functions At an Edge), aimed at allocating VNFs at the CPE nodes (multiple NFVI-PoPs) to minimize end-to-end latency. For this, a general-purpose computer replaces the old proprietary hardware-based CPE. The new CPE operating system is the CentOS 6.4 running a Xen Hypervisor \cite{xen} for network function instantiation as VNFs, and an Open vSwitch (OVS) for VNF interconnection. NetFATE works as the MANO framework, using the NFV Coordinator (C++ software) to manage the VNF life cycle and a POX controller to control the OVSs. Finally, the Orchestration Engine determines how to distribute the VNFs and compose the network services. 

When we consider the use of multiple VIMs without distributed NFV, we usually have a scenario where there are two secondary VIMs, one to manage a data center for virtualization purposes (Cloud domain)
and another to manage a transport network (WAN domain) for end-to-end network services provisioning. As an example, PoC 16 \cite{poc_etsi_16} proposed a multi-domain NFV/SDN architecture intended to provide enterprise services (firewall, IPS/IDS, and load balancer) to remote users across an MPLS-based transport network. This PoC uses an OpenStack (secondary VIM) as cloud orchestrator for VNF instantiation while ensuring end-to-end connectivity and SLAs over the WAN by using OpenFlow with Ryu controller (secondary VIM). The NFV Orchestrator acts as both NFVO and primary VIM, controlling the secondary VIMs to instantiate the end-to-end network services.
\section{SDN-SIDE DESIGN}
\label{sec:sdnside}

This section uses the SDN-side of the taxonomy described in Section \ref{sec:taxonomy} to organize the NFV/SDN solutions regarding the SDN.

\begin{table}[ht]
	\caption{The position of virtual switches in NFV Framework.}
	\label{tab:q343}
	\begin{minipage}{\columnwidth}
		\begin{center}
			\begin{tabular}{p{3cm}p{10cm}}
				\toprule
				\textbf{Position} & \textbf{Studies} \\ \midrule
				
				NFVI & \cite{munoz_integrated_2015,Szabo2015,deng_vnguard:_2015,neves_selfnet_2016,cziva_gnfc:_2015,sonkoly_unifying_2015,cerrato_toward_2015,soares_toward_2015,cziva_container-based_2015,gronsund_solution_2015,soares_cloud4nfv:_2014,vilalta_sdn/nfv_2015,vilalta_transport_2015,vilalta_multitenant_2016,ahmad_new_2016,lucrezia_introducing_2015,mohammadkhan_protocols_2015,akyildiz_wireless_2015,munoz_sdn/nfv_2015,vilalta_multi-tenant_2015,nguyen_slicing_2014,giotis_policy-based_2015,medhat_multi-tenancy_2015,ding_openscaas:_2015,rossem_deploying_2015,lin_extended_2015,callegati_implementing_2015,batalle_implementation_2013,costa-requena_sdn_2015,carella_cross-layer_2015,lombardo_open_2015,lai_rapid_2015,wang_softnet:_2015,mwangama_towards_2015,xia_optical_2015,vestin_qos_2015,schulz-zander_opensdwn:_2015,mamatas_service-aware_2015,poc_etsi_1,poc_etsi_2,poc_etsi_8,poc_etsi_13,poc_etsi_16,poc_etsi_23,poc_etsi_27,poc_etsi_28,poc_etsi_34,poc_etsi_38,7926921,8016637,7945849,7767521,7494060,7718525,7767629,7858117,7980775,7561003,7997431,NEVES2017229} \\
				
				VNF & \cite{haleplidis_forces_2014,haleplidis_towards_2014,neves_selfnet_2016,kim_toward_2015,poc_etsi_13,NEVES2017229}  \\
				
				\bottomrule
			\end{tabular}
		\end{center}
	\end{minipage}
\end{table}

\subsection{Placement of SDN Elements in the NFV Framework} \label{subsec:sdnside_pos}

Table \ref{tab:q343} shows how the studies place virtual switches in the given NFV Framework. The NFVI is the most used as a location for SDN resources. This scenario is a common approach to providing network programmability and flexibility for connectivity and traffic steering among VNFs. 

However, some works also include virtual switches as VNFs. These works intend to provide an SDN-enabled virtual network to different customers. In \cite{haleplidis_forces_2014,haleplidis_towards_2014}, this placement is possible because they work with the Forwarding and Control Element Separation (ForCES) protocol \cite{RFC5810} as SDN Southbound API and consider the Logical Functional Blocks (LFBs) as VNFs (detailed in Section \ref{sub:forces}). On the other hand, Neves et al. \cite{neves_selfnet_2016,NEVES2017229} created an abstraction for deployment of network services through the instantiation of Virtual Network Elements (VNE). VNEs are VNFs running a virtual switch process that perform packet processing (networking services) over the network traffic. VNEs can be distributed in the core or the edge NFVI-PoPs (see Section \ref{sub:mec}).  

Table \ref{tab:q342} shows how the studies positioned the SDN Controllers in the given NFV Framework. According to the \cite{etsi_sdn_in_nfv}, an SDN Controller can run in five (5) places: NFVI, VIM, VNF, OSS/BSS, and can be a  Physical Network Function (PNF). In this work, we did not find references for the last 2 (two) placements. Table \ref{tab:q341} shows how the studies positioned the SDN Applications in the NFV Framework. According to the \cite{etsi_sdn_in_nfv}, SDN Applications can run in five (5) points: as part of a PNF, as part of the VIM, virtualized as a VNF, as part of an Element Manager (EM), and as part of the OSS/BSS. In this work, we did not find references to SDN Applications as part of a PNF or an EM.

\begin{figure}[htbp]
	\centering
	\begin{adjustbox}{width=\textwidth}
		\includegraphics[]{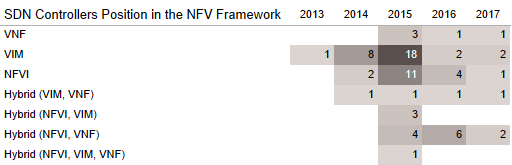}
	\end{adjustbox}
	\caption{Number of studies addressing the position of SDN Controllers in the NFV Framework.}
	\label{fig:cont-pos}
\end{figure}

The VIM is the most used as a position for both SDN Controllers and Applications (see Figure \ref{fig:cont-pos}). The VIM is the best place for these elements because it offers a global view of both NFVI physical and virtual infrastructures and the VNFs. This property allows the implementation of different functionalities, such as VNFs or VNFCs interconnections, network connectivity in the NFVI, and the control of virtual networks as shown in Table \ref{tab:q39}.

However, the NFVI have also been widely adopted as SDN Controller placement. According to \cite{etsi_sdn_in_nfv}, this scenario is a classic case of the SDN controller providing network connectivity in the NFVI. In this work, we consider the SDN controller as a NFVI component when the VIM and its functions are distinguishable, as in the case of OpenStack controlling OpenDaylight. 

\begin{table}[ht]
	\caption{The position of SDN Controllers in the NFV Framework.}
	\label{tab:q342}
	\begin{minipage}{\columnwidth}
		\begin{center}
			\begin{tabular}{p{3cm}p{10cm}}
				\toprule
				\textbf{Position} & \textbf{Studies} \\ \midrule
				
				NFVI & \cite{medhat_multi-tenancy_2015,lucrezia_introducing_2015,carella_cross-layer_2015,rossem_deploying_2015,callegati_implementing_2015,cerrato_toward_2015,soares_toward_2015,soares_cloud4nfv:_2014,munoz_integrated_2015,munoz_sdn/nfv_2015,vilalta_sdn/nfv_2015,vilalta_multi-tenant_2015,vilalta_transport_2015,vilalta_multitenant_2016,Szabo2015,deng_vnguard:_2015,sonkoly_unifying_2015,gronsund_solution_2015,ahmad_new_2016,lin_extended_2015,poc_etsi_8,poc_etsi_21,poc_etsi_28,poc_etsi_34,poc_etsi_38,7926921,7767521,7494060,7718525,7767629,7858117,7980775,7561003,7579021} \\ 
				
				VIM & \cite{haleplidis_forces_2014,Szabo2015,Matias2015,neves_selfnet_2016,cziva_gnfc:_2015,sonkoly_unifying_2015,cziva_container-based_2015,cheng_enabling_2015,mohammadkhan_protocols_2015,shen_vconductor:_2015,akyildiz_wireless_2015,haleplidis_towards_2014,nguyen_slicing_2014,giotis_policy-based_2015,ding_openscaas:_2015,rossem_deploying_2015,batalle_implementation_2013,carella_cross-layer_2015,lombardo_open_2015,lai_rapid_2015,wang_softnet:_2015,mwangama_towards_2015,xia_optical_2015,vestin_qos_2015,saridis_lightness:_2016,basta_sdn_2014,an_virtualization_2014,kyung_software_2015,schulz-zander_opensdwn:_2015,mamatas_service-aware_2015,kim_toward_2015,poc_etsi_1,poc_etsi_2,poc_etsi_23,poc_etsi_27,8016637,7945849,7997431,NEVES2017229} \\ 
				
				VNF & \cite{munoz_integrated_2015,munoz_sdn/nfv_2015,vilalta_sdn/nfv_2015,vilalta_multi-tenant_2015,vilalta_multitenant_2016,neves_selfnet_2016,nguyen_slicing_2014,rossem_deploying_2015,costa-requena_sdn_2015,poc_etsi_13,poc_etsi_16,poc_etsi_26,poc_etsi_27,7767521,7494060,7718525,7767629,7858117,7980775,7561003,7997391,NEVES2017229} \\ 
				
				\bottomrule
			\end{tabular}
		\end{center}
	\end{minipage}
\end{table}

A good example to illustrate SDN Controllers and Applications placement is the work of Rossem et al. (2015) \cite{rossem_deploying_2015}. In that work, the authors have used ESCAPE (see Section \ref{sub:midvirt}) environment to implement a NFV/SDN solution for elastic virtual router provisioning, needed in a VPN service. The main goal is to increase the throughput by load balancing (using Valliant Load Balancing) traffic among multiple virtual switches. In this architecture, the Service Layer receives the VPN requests and define the required VNFs to be instantiated by Orchestration Layer (OL) in an optimal way. There are three VNF types: \textit{Ctrl App}, \textit{OF Ctrl}, and SDN-enabled virtual switches. The \textit{Ctrl App} and \textit{OF Ctrl} are deployed as VMs in OpenStack (Cloud domain), while SDN switches are deployed in an Mininet\footnote{http://mininet.org/} emulator (representing an OpenFlow domain). On the data plane, an elastic router comprises one or more SDN switches. On the control plane, the \textit{SDN Ctrl} manages the topology creation on the top of SDN switches. Moreover, the SDN application \textit{Ctrl App} monitors the SDN flow statistics and triggers topology changes (if needed), adding more or less SDN switches. 

In \cite{rossem_deploying_2015}, a hybrid solution for the positioning of SDN controllers was proposed. In this case, the POX Controller \cite{pox} was placed on VIM to support the creation and management of network services in OpenFlow domains. The OpenDaylight \cite{opendaylight} was placed on NFVI and is used by the OpenStack \cite{openstack} to provide connectivity in the Cloud domain. Finally, the \textit{SDN Ctrl} is a Ryu Controller created as a VNF to coordinate the SDN-enabled virtual network. 

\begin{table}[ht]
	\caption{The position of SDN Applications in the NFV Framework.}
	\label{tab:q341}
	\begin{minipage}{\columnwidth}
		\begin{center}
			\begin{tabular}{p{3cm}p{10cm}}
				\toprule
				\textbf{Position} & \textbf{Studies} \\ \midrule
				
				VIM & \cite{carella_cross-layer_2015,callegati_implementing_2015,neves_selfnet_2016,deng_vnguard:_2015,munoz_integrated_2015,munoz_sdn/nfv_2015,vilalta_sdn/nfv_2015,vilalta_multi-tenant_2015,vilalta_transport_2015,vilalta_multitenant_2016,Szabo2015,cziva_gnfc:_2015,sonkoly_unifying_2015,cerrato_toward_2015,soares_toward_2015,cziva_container-based_2015,soares_cloud4nfv:_2014,lucrezia_introducing_2015,mohammadkhan_protocols_2015,shen_vconductor:_2015,nguyen_slicing_2014,medhat_multi-tenancy_2015,ding_openscaas:_2015,rossem_deploying_2015,lin_extended_2015,lombardo_open_2015,xia_optical_2015,vestin_qos_2015,schulz-zander_opensdwn:_2015,mamatas_service-aware_2015,kim_toward_2015,poc_etsi_1,poc_etsi_2,poc_etsi_8,poc_etsi_16,poc_etsi_27,poc_etsi_28,poc_etsi_34,poc_etsi_38,7926921,8016637,7945849,7767521,7494060,7718525,7767629,7858117,7980775,7561003,7579021,NEVES2017229,7997431} \\ 
				
				VNF & \cite{gronsund_solution_2015,haleplidis_forces_2014,Matias2015,neves_selfnet_2016,ahmad_new_2016,akyildiz_wireless_2015,nguyen_slicing_2014,giotis_policy-based_2015,rossem_deploying_2015,batalle_implementation_2013,costa-requena_sdn_2015,lai_rapid_2015,wang_softnet:_2015,an_virtualization_2014,kyung_software_2015,poc_etsi_13,poc_etsi_21,poc_etsi_23,poc_etsi_26,poc_etsi_27,7997391,NEVES2017229} \\ 
				
				OSS/BSS & \cite{munoz_integrated_2015,munoz_sdn/nfv_2015,vilalta_sdn/nfv_2015,vilalta_multi-tenant_2015,vilalta_transport_2015,vilalta_multitenant_2016,Szabo2015,cheng_enabling_2015,haleplidis_towards_2014,mwangama_towards_2015,xia_optical_2015,saridis_lightness:_2016,basta_sdn_2014,schulz-zander_opensdwn:_2015,mamatas_service-aware_2015,7767521,7494060,7718525,7767629,7858117,7980775,7561003} \\ 
				
				\bottomrule
			\end{tabular}
		\end{center}
	\end{minipage}
\end{table}				

Regarding SDN Applications placement, the \textit{SDN App} also runs as a VNF on top of \textit{SDN Ctrl} (Ryu Controller). For OpenFlow domains, the SDN applications run as a VIM component, on top of POX. Finally, for Cloud domains, the Neutron service (VIM) performs the control of OpenDaylight instance.

The Operations support systems (OSS) and Business Support Systems (BSS) have also been widely adopted as SDN Applications placement. Application at this level enables multiple tenants to control dedicated SDN networks to provide their own services. This scenario is common in works that propose solutions for 5G Cellular Networks \cite{neves_selfnet_2016,NEVES2017229,mwangama_towards_2015}. Examples of SDN application placement as OSS/BSS management task are the works of Munoz and Vilalta \cite{munoz_integrated_2015,munoz_sdn/nfv_2015,vilalta_sdn/nfv_2015,vilalta_multi-tenant_2015,vilalta_multitenant_2016,7767521,7494060,7718525,7767629,7858117,7980775,7561003}. In those works, a tenant SDN Controller runs as a VNF to control an underlying VTN. The end-users or service provider operators (components of OSS/BSS element) have direct access to this controller and can implement customized applications for VTN control and management.

\subsection{SDN Controller Functions in the NFV Framework} \label{subsec:sdnside_fun}

Table \ref{tab:q39} shows the possible functions for SDN Controllers when applied in the NFV Framework. Below, we describe these functions:

\begin{description}
	\item[Interconnecting VNFs/VNFCs:] The SDN Controller might be used to connect and manage the traffic between VNFs and/or VNFCs to enable Network Services, by creating Service Function Chaining (SFC). As seen in Section \ref{sub:nfv}, a VNF might be composed of several VNFCs.	
	\item[Network Connectivity in the NFVI:] The SDN Controller is used to provide L2/L3 connectivity among end-to-end devices;
	\item[Control of Virtual Networks:] The SDN Controller might be responsible for creating and managing virtual networks for different customers.
\end{description}

\begin{figure}[htbp]
	\centering
	\begin{adjustbox}{width=\textwidth}
		\includegraphics[]{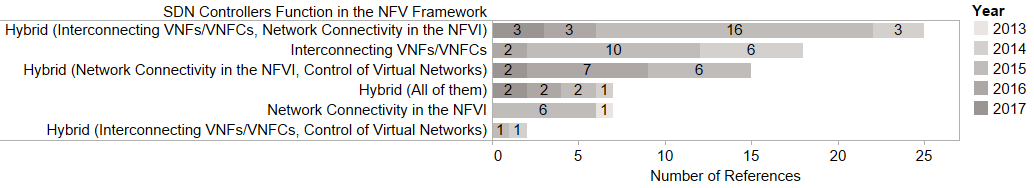}
	\end{adjustbox}
	\caption{Number of studies addressing SDN Controller functions.}
	\label{fig:cont-func}
\end{figure}

\begin{table}[ht]
	\caption{List of possible functions for SDN Controllers when applied in NFV Framework.}
	\label{tab:q39}
	\begin{minipage}{\columnwidth}
		\begin{center}
			\begin{tabular}{p{5cm}p{8cm}}
			\toprule
			\textbf{Function} & \textbf{Studies} \\ \midrule
			
			Interconnecting VNFs/VNFCs & \cite{carella_cross-layer_2015,haleplidis_forces_2014,Szabo2015,neves_selfnet_2016,cziva_gnfc:_2015,sonkoly_unifying_2015,cerrato_toward_2015,soares_toward_2015,cziva_container-based_2015,gronsund_solution_2015,soares_cloud4nfv:_2014,cheng_enabling_2015,ahmad_new_2016,lucrezia_introducing_2015,mohammadkhan_protocols_2015,haleplidis_towards_2014,nguyen_slicing_2014,medhat_multi-tenancy_2015,ding_openscaas:_2015,rossem_deploying_2015,lin_extended_2015,callegati_implementing_2015,lombardo_open_2015,lai_rapid_2015,wang_softnet:_2015,mwangama_towards_2015,xia_optical_2015,basta_sdn_2014,an_virtualization_2014,kyung_software_2015,schulz-zander_opensdwn:_2015,mamatas_service-aware_2015,kim_toward_2015,poc_etsi_1,poc_etsi_2,poc_etsi_8,poc_etsi_13,poc_etsi_16,poc_etsi_21,poc_etsi_23,poc_etsi_26,poc_etsi_27,poc_etsi_28,poc_etsi_34,poc_etsi_38,7926921,8016637,7945849,7997431,7997391,7579021,NEVES2017229} \\ 
			
			Network Connectivity in the NFVI & \cite{gronsund_solution_2015,vestin_qos_2015,lucrezia_introducing_2015,callegati_implementing_2015,rossem_deploying_2015,cerrato_toward_2015,soares_toward_2015,soares_cloud4nfv:_2014,deng_vnguard:_2015,munoz_integrated_2015,munoz_sdn/nfv_2015,vilalta_sdn/nfv_2015,vilalta_multi-tenant_2015,vilalta_transport_2015,vilalta_multitenant_2016,Szabo2015,Matias2015,neves_selfnet_2016,sonkoly_unifying_2015,ahmad_new_2016,shen_vconductor:_2015,akyildiz_wireless_2015,giotis_policy-based_2015,medhat_multi-tenancy_2015,batalle_implementation_2013,costa-requena_sdn_2015,carella_cross-layer_2015,lombardo_open_2015,saridis_lightness:_2016,basta_sdn_2014,kyung_software_2015,schulz-zander_opensdwn:_2015,kim_toward_2015,poc_etsi_1,poc_etsi_8,poc_etsi_16,poc_etsi_21,poc_etsi_26,poc_etsi_27,poc_etsi_34,poc_etsi_38,7926921,8016637,7945849,7767521,7494060,7718525,7767629,7858117,7980775,7561003,7997391,7579021,NEVES2017229} \\ 
			
			Control of Virtual Networks &  \cite{vestin_qos_2015,neves_selfnet_2016,munoz_integrated_2015,munoz_sdn/nfv_2015,vilalta_sdn/nfv_2015,vilalta_multi-tenant_2015,vilalta_multitenant_2016,akyildiz_wireless_2015,nguyen_slicing_2014,mwangama_towards_2015,saridis_lightness:_2016,kyung_software_2015,schulz-zander_opensdwn:_2015,poc_etsi_1,poc_etsi_38,8016637,7767521,7494060,7718525,7767629,7858117,7980775,7561003,NEVES2017229} \\
			
			\bottomrule
			\end{tabular}
		\end{center}
	\end{minipage}
\end{table}		

It is worth emphasizing that interconnecting VNFs is the main objective for using SDN in the NFV Framework (see Figure \ref{fig:cont-func}), mainly for automation and monitoring of SFC deployments. The works of Cziva el al.
\cite{cziva_gnfc:_2015,cziva_container-based_2015,7945849} used an OpenDaylight Controller to create OpenFlow rules in Open vSwitch instances and OpenFlow switches, intending to provide interconnectivity and traffic steering for network services (set of linked VNFs). According to \cite{7243304}, there are some problems regarding Interconnecting VNFs, such as network function placement, survivability of VNFs, and dynamic service function chaining (elasticity).

As an example of SDN Controllers performing Network Connectivity in the NFVI and Control of Virtual Networks, Vestin et al. (2015) \cite{vestin_qos_2015} used OpenDaylight Controller as VIM to manage the communication among physical APs and Virtual Access Points (VAP) (see Section \ref{sub:wlan}). In this case, the controller provides Network connectivity in the NFVI by connecting physical AP and Cloud infrastructure, and Control of Virtual Networks by using OpenFlow to connect a mobile client with its respective VAP. 

\subsection{The Use of Multiple SDN Controllers} \label{subsec:sdnside_mul}

Table \ref{tab:q38} shows the main objectives for implementing Multiple SDN Controllers in the NFV Framework. As well as the use of Multiple VIMs (see Section \ref{sec:nfvside}), multiple SDN controllers are organized hierarchically to provide the following objectives. 

\begin{description}
	\item[Distributed Performance:] When the VNFs have to be distributed to the chosen location, they are interconnected by location-specific SDN Controllers;
	\item[Scalability:] Multiple controllers manage an NFVI-PoP infrastructure;
	\item[Reliability:] Fault tolerance, disaster recovery, and full isolation management.  
	\item[Administrative Domains Interaction:] Communication management of different scenarios (e.g., Cloud and WAN) using different SDN Controllers integrated hierarchically in a unique platform; 
	\item[Network as a Service (NaaS) Management:] The concept of NaaS is related to the provision of virtualized network services to customers with different requirements \cite{Mustafa20151}. This function includes network virtualization (Network Slicing).
\end{description}

\begin{table}[ht]
	\caption{List of objectives for use Multiple SDN Controllers.}
	\label{tab:q38}
	\begin{minipage}{\columnwidth}
		\begin{center}
			\begin{tabular}{p{5cm}p{8cm}}
				\toprule
				\textbf{Objective} & \textbf{Studies} \\ \midrule
				
				Distributed Performance & \cite{poc_etsi_34} \\  
				
				Scalability & \cite{callegati_implementing_2015}\\ 
				
				Reliability & \cite{munoz_integrated_2015,munoz_sdn/nfv_2015,vilalta_sdn/nfv_2015,vilalta_multi-tenant_2015,vilalta_multitenant_2016,7767521,7494060,7718525,7767629,7858117,7980775,7561003} \\
				
				Administrative Domains Interaction & \cite{rossem_deploying_2015,neves_selfnet_2016,munoz_integrated_2015,munoz_sdn/nfv_2015,vilalta_sdn/nfv_2015,vilalta_multi-tenant_2015,vilalta_transport_2015,vilalta_multitenant_2016,Szabo2015,sonkoly_unifying_2015,batalle_implementation_2013,carella_cross-layer_2015,kyung_software_2015,poc_etsi_1,poc_etsi_16,7926921,7767521,7494060,7718525,7767629,7858117,7980775,7561003,7579021,NEVES2017229}\\ 
				
				NaaS Management & \cite{neves_selfnet_2016,munoz_integrated_2015,munoz_sdn/nfv_2015,vilalta_sdn/nfv_2015,vilalta_multi-tenant_2015,vilalta_multitenant_2016,Szabo2015,cerrato_toward_2015,akyildiz_wireless_2015,rossem_deploying_2015,poc_etsi_1,poc_etsi_16,7926921,cziva_gnfc:_2015,cziva_container-based_2015,7945849,7767521,7494060,7718525,7767629,7858117,7980775,7561003,NEVES2017229}\\ 
				
				\bottomrule
			\end{tabular}
		\end{center}
	\end{minipage}
\end{table}	

In PoC 34 \cite{poc_etsi_34}, the developers implement Distributed Performance by putting EPC Virtual Functions to different NFVI-PoPs, each NFVI-PoP managed by its own SDN Controller (OpenDaylight). These controllers communicate with each other to allow the programming of GPRS Tunneling Protocol (GTP) tunnels interconnecting VNFs placed in different locations.

Scalability is handled in Callegati et al. (2015) \cite{callegati_implementing_2015}. In this work, the authors used two SDN controllers to manage the virtual networks in an OpenStack-based Cloud environment. They create OpenFlow rules using a Neutron Open vSwitch Agent for connectivity in NFVI and interconnection of VNFs, and monitor the throughput of OpenFlow rules using a POX \cite{pox} controller for traffic steering mechanisms.  

Regarding Reliability, the works of Munoz et al. (2015) \cite{munoz_integrated_2015,munoz_sdn/nfv_2015} used NFV and Cloud to virtualize tenant SDN Controllers to control the underlying VTNs. The authors pointed out that a clear advantage of using cloud virtualization for SDN controllers is the reliability achieved with the lack of hardware maintenance downtime and the decreasing recovery time.

However, it is worth noting that Administrative Domains Interaction and NaaS Management are the main objectives for using multiple SDN controllers in the NFV Framework. As an example, Rossem et al. (2015) \cite{rossem_deploying_2015} used three SDN controllers in their NFV/SDN architecture to provide elastic virtual router provisioning in a multi-domain scenario (described in Section \ref{subsec:sdnside_pos}). For Administrative Domains Interaction, the authors used OpenDaylight and POX controllers to provide an SDN-enabled virtual network on top of a Cloud (OpenStack-based) and a WAN domain (OpenFlow-based), respectively. Finally, for NaaS Management, the authors instantiate SDN Controllers (Ryu) as VNFs to control the underlying virtual networks.

\section{AUXILIARY TOOLS}
\label{sec:tools-nfv-sdn}

This section presents the tools used in the studies carried out to implement NFV/SDN architectures. The primary focus is on free or open-source tools, which are thus freely available for future implementations.

\subsection{VNF Instantiation}

The following open source tools have been used in NFVI to enable virtualization of network functions.  

\begin{description}
	\item[Kernel Virtual Machine (KVM) \cite{kvm}:] KVM enables Linux Kernel as hardware-assisted virtualization hypervisor on x86 and x86-64 systems equipped with virtualization extensions (Intel VT or AMD-V). KVM allows the running of multiple virtual machines with unmodified Unix-based systems (e.g., Linux or NetBSD) and Windows images. All studies that applied OpenStack as VIM (see \ref{sub:openstack}) used KVM because this is the OpenStack default hypervisor.    
	\item[Xen Hypervisor \cite{xen}:] Xen is a hypervisor that allows the creation of multiple virtual machines, using paravirtualization capabilities in x86, x86-64, ARM, and PowerPC architectures. By working with paravirtualization, Xen uses modified images (e.g., Linux or Windows) to provide fast execution of virtual machines. As seen in Section \ref{sec:nfvside}, Lombardo et al. (2015) used Xen to enable the rapid deployment of new VNFs on vCPE nodes \cite{lombardo_open_2015}.
	\item[ClickOS \cite{clickos}:] ClickOS is a high-performance Xen-based software platform that enables VNF development using a Click modular router software running on top of MiniOS. It allows the creation of small VMs (5MB) with a fast bootloader (30 milliseconds). As seen in Section \ref{sub:midvirt}, Deng et al. (2015) used ClickOS for its virtual firewall development \cite{deng_vnguard:_2015}.
	\item[Docker Engine \cite{docker}:] Docker uses container-based virtualization to create multiple isolated containers that run natively on Windows, Linux and MacOS systems. A container is just a package that contains a set of libraries needed to run the hosted applications. Unlike virtual machines, containers do not have a dedicated operating system, sharing the features of the host operating system. As a consequence, containers are faster and consume less computational resources, enabling improvements in VNF provisioning time (including up/down/update), runtime performance (e.g., throughput), and in the number of VNFs (hundreds) that a commodity computing devices can host \cite{natarajan-nfvrg-containers-for-nfv-03}. As seen in Section \ref{sub:midvirt}, Cziva et al. \cite{cziva_gnfc:_2015,cziva_container-based_2015,7945849} used Docker Engine to provide fast deployment of new virtual middleboxes with less resource utilization.   
	\item[Data Plane Development Kit (DPDK) \cite{dpdk}:] DPDK allows high-performance packet processing through a set of data libraries and drivers for network interface controlling. The VNF developer can use such libraries to create network functions with high throughput. These functions run as an isolated process and communicate through xDPd OpenFlow Switches (detailed in Section \ref{sub:virtualswitches}). \cite{cerrato_toward_2015} used DPDK processes in its integrated node to create VNFs with fast packet processing in equipment with limited resources.    
\end{description}

For more information on paravirtualization and hardware-assisted virtualization, we refer the reader to the following white paper from VMWare\footnote{http://www.vmware.com/techpapers/2007/understanding-full-virtualization-paravirtualizat-1008.html}: ``Understanding Full Virtualization, Paravirtualization, and Hardware Assist''. For more information on container-based virtualization, please read the Docker definition\footnote{https://www.docker.com/what-container}.



\subsection{OpenStack Cloud Platform}\label{sub:openstack}

Table \ref{tab:openstack} lists all studies that used OpenStack as part of their solutions. All of them used OpenStack as VIM for cloud domain management, seeking to provide a platform for VNF virtualization. 

Considered one of the most significant open source projects, the OpenStack is a Cloud Platform that emerged in 2010 through an initiative of Rackspace Hosting\footnote{Rackspace website: https://www.rackspace.com/} and NASA. Its structure was based on the NASA platform Nebula and the Rackspace cloud file system. Since 2012, the project has been managed by OpenStack Foundation.

\begin{table}[ht]
	\caption{List of works using OpenStack.}
	\label{tab:openstack}
	\begin{minipage}{\columnwidth}
		\begin{center}
			\begin{tabular}{p{3cm}p{10cm}}
				\toprule
				\textbf{Tool} & \textbf{Studies} \\ \midrule
				OpenStack & \cite{munoz_integrated_2015,munoz_sdn/nfv_2015,deng_vnguard:_2015,cerrato_toward_2015,soares_toward_2015,gronsund_solution_2015,soares_cloud4nfv:_2014,vilalta_sdn/nfv_2015,vilalta_transport_2015,vilalta_multitenant_2016,lucrezia_introducing_2015,shen_vconductor:_2015,vilalta_multi-tenant_2015,medhat_multi-tenancy_2015,rossem_deploying_2015,callegati_implementing_2015,costa-requena_sdn_2015,carella_cross-layer_2015,mwangama_towards_2015,poc_etsi_1,poc_etsi_8,poc_etsi_13,poc_etsi_28,poc_etsi_34,7767521,7494060,7718525,7767629,7858117,7980775,7561003} \\
				\bottomrule
			\end{tabular}
		\end{center}
	\end{minipage}
\end{table}

The OpenStack Cloud Platform provides an Infrastructure as a Service (IaaS) solution through a set of related services, written in Python. Below we describe the main services.

\begin{description}
	\item[KeyStone:] Represents the Identity Service. It provides functionalities such as authentication, authorization, and service catalog.
	\item[Nova:] Represents the Computing Service. It is responsible for managing the lifecycle of virtual machine instances. Currently, it provides support for different hypervisors: Xen, XenServer/XCP, QEMU, KVM, UML, VMware, vSphere, and Hyper-V.
	\item[Neutron:] Represents the Networking Service. This service provides connectivity to the interfaces of the VMs. It has a flexible API that allows the construction of complex networks: flat and shared networks, VLANs, VXLAN, GRE, DHCP, IPv6, and SDN, for example.
	\item[Glance:] Represents the Image Service. It stores and retrieves disk images of virtual machines. It also stores metadata of these images.
	\item[Swift:] Represents the Object Storage Service. It stores and retrieves unstructured data objects. It works with data replication to provide a highly tolerant and scalable architecture.
	\item[Cinder:] Represents the Block Storage Service. It provides a persistent storage block for running instances.
\end{description}

\subsection{Virtual Switches}\label{sub:virtualswitches}

The \textbf{Open vSwitch (OVS)} \cite{ovs} was the most adopted virtual switch. Currently maintained by the Linux Foundation\footnote{https://www.linuxfoundation.org/}, OVS aims to automate network tasks. Therefore, it supports many protocols, such as OpenFlow (versions 1.0 \cite{onfspecopenflow10} and 1.3 \cite{onfspecopenflow13}), OVSDB \cite{rfc7047}, NetFlow, sFlow, IPFIX, and the like. It also provides many additional features, such as VLAN isolation, traffic filtering, traffic queuing, and traffic shaping. Furthermore, the OVS has also been integrated into popular cloud platforms including oVirt, OpenNebula, and OpenStack. 

GNF (Glasgow Network Functions) \cite{cziva_gnfc:_2015,cziva_container-based_2015,7945849} proposes an NFV Framework for public clouds and uses OVS to provide the data plane required to interconnect network functions (Docker container-based) and connect them to arbitrary services. In Vestin and Kassler (2015), an extended OVS was installed in 802.11 Access Points to provide Virtual Access Points with QoS enforcement. The NetFATE (Network Functions At the Edge) platform \cite{lombardo_open_2015} uses the OVS to implement NFV services at the edge of a Telco network, providing data plane in CPE nodes.

Carella et al. (2015) \cite{carella_cross-layer_2015} adopted the \textbf{OpenSDNCore Switches (OSCS)} \cite{opensdncore} to provide resource reservation through the creation of flow entries and priority queues in a network layer. The OSCS is a software implementation (in C language) of an extended OpenFlow 1.4 switch with specific telecom oriented extensions supporting many features, such as OpenFlow 1.4, GPRS Tunneling Protocol (GTP), Generic Routing Encapsulation (GRE), asynchronous metrics (statistics), traffic shaping, and topology learning.

Finally, the \textbf{eXtensible Datapath daemon (xDPd)} \cite{dpdk} provides a framework for building multiples high-performance OpenFlow datapath elements, called Logical Switch Instances (LSIs). xDPd is written in C/C++ and supports the following platforms: GNU/Linux amd64/x86 user-space, GNU/Linux Intel's DPDK accelerated driver, NetFGPA-10G (netfpga10g), Broadcom Triumph2 (bcm), and Octeon network processors. xDPs support multiple OpenFlow versions (1.0, 1.2, and 1.3). Cerrato et al. (2015) used xDPd to create multiples LSIs (one for each customer) in the integrated node equipment with limited resources that represents a CPE \cite{cerrato_toward_2015} (as described in Section \ref{sub:vcpe}).   

\subsection{SDN Controllers}

\begin{table}[ht]
	\caption{List of works grouped by the used SDN Controllers.}
	\label{tab:controllers}
	\begin{minipage}{\columnwidth}
		\begin{center}
			\begin{tabular}{p{3cm}p{10cm}}
				\toprule
				\textbf{SDN Controller} & \textbf{Studies} \\ \midrule
				
				OpenDaylight & \cite{munoz_integrated_2015,cziva_gnfc:_2015,sonkoly_unifying_2015,cerrato_toward_2015,soares_toward_2015,cziva_container-based_2015,gronsund_solution_2015,soares_cloud4nfv:_2014,vilalta_sdn/nfv_2015,vilalta_transport_2015,vilalta_multitenant_2016,lucrezia_introducing_2015,shen_vconductor:_2015,munoz_sdn/nfv_2015,vilalta_multi-tenant_2015,medhat_multi-tenancy_2015,vestin_qos_2015,saridis_lightness:_2016,poc_etsi_34,7945849,7767521,7494060,7718525,7767629,7858117,7980775,7561003}\\ 
				ONOS & \cite{8016637,7997431} \\
				
                Floodlight & \cite{munoz_integrated_2015,cheng_enabling_2015,vilalta_sdn/nfv_2015,munoz_sdn/nfv_2015,vilalta_multi-tenant_2015,batalle_implementation_2013,schulz-zander_opensdwn:_2015,poc_etsi_21}\\ 
				
				Ryu & \cite{rossem_deploying_2015,lin_extended_2015,costa-requena_sdn_2015,carella_cross-layer_2015,lai_rapid_2015,poc_etsi_2,poc_etsi_16,poc_etsi_26}\\ 
				
				POX & \cite{sonkoly_unifying_2015,rossem_deploying_2015,callegati_implementing_2015,lombardo_open_2015}\\ 
				\bottomrule
			\end{tabular}
		\end{center}
	\end{minipage}
\end{table}

Table \ref{tab:controllers} lists all SDN controllers used in implementations of NFV/SDN architectures. As seen in Section \ref{subsec:sdnside_pos}, such controllers have been used as NFVI, VIM or VNF components. Below, we present a brief description of them.

\begin{description}
	\item[OpenDaylight (ODL) \cite{opendaylight}:] The ODL is a modular SDN open source platform maintained by The Linux Foundation\footnote{The Linux Foundation website: www.linuxfoundation.org/}. Written in Java, the ODL aims to accelerate the development of solutions for SDN and NFV in production environments. ODL offers plugins that support different SDN Southbound API, such as OpenFlow (1.0 and 1.3), LISP, NETCONF, and OVSDB. Currently, the newest version of the ODL is the Boron, released in December 2016.
	\item[Open Network Operating System (ONOS) \cite{onos}:] ONOS comprises an open source SDN Controller focused on the construction of NFV/SDN solutions. Like ODL, ONOS was developed in Java on top of the Apache Karaf OSGi container and provides the following features: i) a GUI for the view the network state, support different SDN Southbound API, such as OpenFlow, NETCONF, and OpenConfig; ii) northbound abstractions to simplify the creation of intent-based virtualized networks; iii) high availability and scalability support (e.g., cluster of ONOS instances).    
    \item[Floodlight \cite{floodlight}:] Floodlight is a SDN Controller written in Java that supports the following OpenFlow versions: 1.0, 1.1, 1.2, 1.3, and 1.4. It is Apache-licensed and supported by engineers and developers from Big Switch Networks. In addition to OpenFlow controller, Floodlight provides a set of internal SDN applications (e.g., firewall and load balancing) and a REST API for development of external applications.
	\item[Ryu \cite{ryu}:] Ryu is an open source framework (Apache 2.0 licensed) created by NTT and written in Python. Ryu supports several southbound interfaces, such as OpenFlow, OF-config, and NETCONF. Regarding OpenFlow, Ryu supports the following versions: 1.0, 1.2, 1.3, 1.4, and 1.5. A REST API is available to be used for external SDN applications. Currently, Ryu is fully integrated into Neutron (OpenStack Networking Service).
	\item[POX \cite{pox}:] Developed at Stanford University, POX was one of the first open source developed SDN controllers. Written in Python, POX only supports OpenFlow version 1.0. It is currently a discontinued project.
\end{description}

The adoption of both ODL and OpenStack to compose NFV/SDN solutions is emerging, mainly due to the soft integration of these tools. The Neutron service uses the Module Layer 2 (ML2) Plugin \cite{openstack} to provide networking services in a Cloud. The ML2 might control an ODL instance using the Neutron API, a REST API provided by ODL. It is worth mentioning that some new NFV/SDN architectures \cite{8016637,7997431} with the focus only on SDN solutions \cite{7502474,7968339} have adopted the ONOS SDN Controller. This controller has gained momentum, and it is currently the main competitor of ODL. ONOS is the official distribution of the Open Network Foundation (ONF) along with the Open Networking Lab (ON.Lab). Furthermore, essential projects started as an ONOS use case, such as Central Office Re-architected as a Datacenter (CORD). CORD combines NFV and SDN technologies to create a general-purpose platform that is capable of delivering a broad range of innovative services targeting network operators, from access services (e.g., Fiber-to-the-Home) to general cloud services (SaaS) \cite{cord}. In addition to ONOS, CORD supports other open source tools such as OpenStack, Docker, etc. Major players such as AT\&T, Google, Cisco, NEC, Nokia, Fujitsu, Intel, SK Telecom, Verizon, China Unicom and NTT Communications are already supporting CORD.

\subsection{SDN Southbound Interfaces}

Table \ref{tab:southbound} lists all the interfaces used as SDN Southbound API in the works. Such APIs are part of two SDN standards: OpenFlow and ForCES. These standards, as well as their interfaces to access network devices, are described below.

\begin{table}[ht]
	\caption{List of works grouped by SDN Southbound API used.}
	\label{tab:southbound}
	\begin{minipage}{\columnwidth}
		\begin{center}
			\begin{tabular}{p{5cm}p{8cm}}
				\toprule
				\textbf{SDN Southbound API} & \textbf{Studies} \\ \midrule 
				ForCES Protocol & \cite{haleplidis_forces_2014,haleplidis_towards_2014}\\ 
				
				OpenFlow Protocol & \cite{cziva_gnfc:_2015,cziva_container-based_2015,deng_vnguard:_2015,Matias2015,cheng_enabling_2015,ahmad_new_2016,mohammadkhan_protocols_2015,akyildiz_wireless_2015,nguyen_slicing_2014,giotis_policy-based_2015,batalle_implementation_2013,costa-requena_sdn_2015,carella_cross-layer_2015,lai_rapid_2015,xia_optical_2015,an_virtualization_2014,kyung_software_2015,poc_etsi_2,poc_etsi_21,poc_etsi_23,poc_etsi_27,poc_etsi_38,sonkoly_unifying_2015,munoz_sdn/nfv_2015,vilalta_sdn/nfv_2015,vilalta_multi-tenant_2015,vilalta_multitenant_2016,rossem_deploying_2015,callegati_implementing_2015,lombardo_open_2015,saridis_lightness:_2016,ding_openscaas:_2015,lin_extended_2015,vestin_qos_2015,schulz-zander_opensdwn:_2015,poc_etsi_8,poc_etsi_13,poc_etsi_16,poc_etsi_26,poc_etsi_28,8016637,7945849,7767521,7494060,7718525,7767629,7858117,7980775,7561003,7997431} \\ 
				
				OVSDB Management Protocol & \cite{cziva_gnfc:_2015,cziva_container-based_2015,munoz_integrated_2015,munoz_sdn/nfv_2015,cerrato_toward_2015,soares_toward_2015,gronsund_solution_2015,soares_cloud4nfv:_2014,vilalta_sdn/nfv_2015,vilalta_transport_2015,vilalta_multi-tenant_2015,vilalta_multitenant_2016,lucrezia_introducing_2015,shen_vconductor:_2015,medhat_multi-tenancy_2015,poc_etsi_1,callegati_implementing_2015,poc_etsi_34,7945849,7767521,7494060,7718525,7767629,7858117,7980775,7561003} \\  
				
				\bottomrule
			\end{tabular}
		\end{center}
	\end{minipage}
\end{table}

\subsubsection{\textbf{OpenFlow}}\label{sub:openflow}

The development of OpenFlow began in 2007, and its evolution has received contributions from both academia and industry. Designed originally by researchers at Stanford University and the University of California at Berkeley, this standard has been maintained by the Open Networking Foundation (ONF) \cite{Kreutz201414}. 

The ONF\footnote{Open Networking Foundation website: https://www.opennetworking.org/index.php} is an organization committed to the development and dissemination of SDN. For this, ONF is responsible for the creation of open standards for SDN, such as the OpenFlow specifications. OpenFlow is frequently updated, thus adding new features from version 1.0 \cite{onfspecopenflow10} to 1.3 \cite{onfspecopenflow13}. Multiple flow tables, group and meter tables, and MPLS support are some of the advances since version 1.1. 

The OpenFlow comprises three main elements, namely switches, controllers, and protocols (Southbound API). The OpenFlow Switches are responsible for the data packet forwarding (i.e., data plane) according to the rules created and maintained by the Controller \cite{onfspecopenflow13}. Multiple Flow Tables store these rules. The controller is the main component of OpenFlow. In architectural terms, the controller supports the network applications, determining the rules to be stored and applied by switches. There are several OpenFlow controllers available, such as OpenDaylight, Floodlight, ONOS, Ryu, POX, and NOX. 

The controller can use two types of interfaces to create OpenFlow rules on switches: the OpenFlow Protocol and the Open vSwitch Database (OVSDB) Management Protocol. Both interfaces are described below.

\begin{description}
	\item[OpenFlow Protocol:] Defined in OpenFlow Specifications \cite{onfspecopenflow13}, this protocol uses a secure and encrypted channel (TCP/TLS) for performing the management of switches. It includes a set of messages for different situations: establishment and configuration of the management channel (e.g., Hello, Echo Request/Reply, Features Request/Reply, Set-Config), receiving (Packet-in) and redirecting (Packet-out) data packets, and OpenFlow rules management (Flow-mod). The OpenFlow protocol can coordinate all OpenFlow-enabled switches.
	\item[OVSDB Management Protocol:] Defined in RFC 7047 \cite{rfc7047}, this protocol uses a set of operations available in Open vSwitch (programmatic extension) to manage OpenFlow rules. Such operations allow insertion, updating, and deletion of forwarding rules directly into the Open vSwitch Database. As a consequence, the OVSDB Management Protocol is limited to use in virtual switches based on Open vSwitch.
\end{description}

As an example of using these two protocols, Callegati et al. (2015) \cite{callegati_implementing_2015} proposed a solution to deploy multi-tenant service function chaining of edge network functions, using OpenStack. For this, the authors used an OVSDB management protocol (via Neutron Open vSwitch Agent) to provide connectivity to VNFs and an OpenFlow protocol (via POX controller) to monitor the throughput of OpenFlow rules.

\subsubsection{\textbf{Forwarding and Control Element Separation (ForCES)}} \label{sub:forces}

ForCES is an SDN standard defined by the Internet Engineering Task Force (IETF) \cite{RFC5810}. In ForCES, the plane separation occurs by dividing the Network Elements (NE) into two entities: Forwarding Elements (FE) and the Control Elements (CE). 

FEs represent the Data Plane and comprise both physical and virtual switches. ForCES models FEs by defining one or many Logical Functional Blocks (LFBs) classes, realized by an XML-based modeling language. An LFB comprises input and output ports and acts as a packet processing resource performing different functions, such as filtering, classification, and measurement. Multiple LFB instances in the same FE can be connected in a directed graph to create a network service.

Each LFB provides operational parameters, capabilities, and events to a CE that acts as an SDN Controller. CE uses the \textbf{ForCES protocol} as a Southbound API to perform the per-LFB controlling.

The works of Haleplidis et al. (2014) \cite{haleplidis_forces_2014,haleplidis_towards_2014} proposed an NFV/SDN architecture using a ForCES standard. In this solution, the NFVI contains an LFB hypervisor that allows the creation of FE/LFBs as VNFs and CEs as EMS entities. The NFVI also provides LFBs acting as virtual switches to provide interconnection between these VNFs. In \cite{haleplidis_forces_2014}, a PoC has been proposed to evaluate this architecture when applied for virtualization of 4G Evolved Packet Core (EPC) components. 

\subsection{SDN Northbound Interfaces}

Regarding Northbound Interface (NBI), all studies that used this type of interface (see Table \ref{tab:nbi}) chose to work with REST API \cite{wsdl2}. According to \cite{7378522}, the REST API has become a prevalent choice for the NBI in SDN, because it is highly extensible and maintainable for managing services from both data and control planes. HTTP \cite{Fielding:1999:HTP:RFC2616} is usually adopted as a protocol for the communication between REST services (called web services). Procera \cite{procera} and Frenetic \cite{frenetic} would be alternatives for NBIs, but they run on top of a single OpenFlow Controller, and they are not so extensible as RESTful interfaces \cite{7378522}.

\begin{table}[ht]
	\caption{List of works using REST API as NBI.}
	\label{tab:nbi}
	\begin{minipage}{\columnwidth}
		\begin{center}
			\begin{tabular}{p{5cm}p{8cm}}
				\toprule
				\textbf{SDN Northbound API} & \textbf{Studies} \\ \midrule
				REST API & \cite{munoz_integrated_2015,cerrato_toward_2015,neves_selfnet_2016,cziva_gnfc:_2015,soares_toward_2015,cziva_container-based_2015,gronsund_solution_2015,soares_cloud4nfv:_2014,vilalta_sdn/nfv_2015,vilalta_transport_2015,vilalta_multitenant_2016,lucrezia_introducing_2015,shen_vconductor:_2015,munoz_sdn/nfv_2015,vilalta_multi-tenant_2015,nguyen_slicing_2014,medhat_multi-tenancy_2015,batalle_implementation_2013,carella_cross-layer_2015,saridis_lightness:_2016,poc_etsi_1,poc_etsi_2,poc_etsi_8,poc_etsi_16,poc_etsi_26,poc_etsi_34,7945849,7767521,7494060,7718525,7767629,7858117,7980775,7561003,7997431,NEVES2017229}\\
				\bottomrule
			\end{tabular}
		\end{center}
	\end{minipage}
\end{table}

RESTful interfaces are available in the main SDN Controllers, such as OpenDaylight, Floodlight, ONOS, and Ryu. 

\subsection{Vendor-specific Tools}

It is also important to note that the Vendor-specific solutions are only used in some PoCs from ETSI NFV ISG, for all NFV components.

Some of the features used are: HP \cite{poc_etsi_23,poc_etsi_27,poc_etsi_34,poc_etsi_38} and Huawei \cite{poc_etsi_28} NFV Orchestrators (NFVO); ZTE \cite{poc_etsi_27}, Riverbed \cite{poc_etsi_28}, Samsung \cite{poc_etsi_23}, Telcoware \cite{poc_etsi_23}, HP \cite{poc_etsi_34} and F5 \cite{poc_etsi_38} VNF Managers (VNFM); and Samsung \cite{poc_etsi_23}, Telcoware \cite{poc_etsi_23}, HP \cite{poc_etsi_27,poc_etsi_38} and ZTE \cite{poc_etsi_27} Virtual Infrastructure Managers (VIM).
\section{LESSONS LEARNED}
\label{sec:lessons}

This section summarizes our view from the literature review described in the previous sections (Sections \ref{sec:app-nfv-sdn}, \ref{sec:nfvside}, \ref{sec:sdnside}, and \ref{sec:tools-nfv-sdn}) and points out some trends for the design and implementation of NFV/SDN architectures. 

Cloud Computing is the dominant scenario for implementing NFV/SDN solutions (72\% of the studies found). According to \cite{7243304}, Cloud Computing and Software-Defined Networking (SDN) are two concepts closely related to NFV. Most of the proposed NFV solutions have been implemented and tested in cloud-based environments. It has been the primary choice for the creation of NFV infrastructures (NFVI) mainly due to its flexibility, rapid deployment of new services, and inherent elasticity. The VNFs of a specific SFC are deployed as functions in dedicated Virtual Machines (VMs), which can be instantiated on devices placed in different geographic locations. Cloud Computing allows NFV/SDN solutions to provide better services for users by simplifying the provision of network services and enabling the quick deployment, management, and optimization of physical infrastructure dynamically, using resource virtualization mechanisms. 

At the SDN-side, the SDN elements have been placed in different points of the NFV framework. It is clear that SDN Switches are most present in NFVI, whereas SDN Controllers are deployed in VIM and at the NFVI. Also, SDN Applications are usually placed in VIMs. The VIM is most used as a position for both SDN Controllers and Applications. The VIM seems to be the best place for these elements because it offers a global view of both NFVI physical and virtual infrastructures and the VNFs. It allows the implementation of different functionalities, such as VNFs or VNFCs interconnections, network connectivity in the NFVI, and the control of virtual networks.

Another important observation is that the most used elements in the VIM component are the OpenStack \cite{openstack} and the SDN Controller OpenDaylight (ODL) \cite{opendaylight} due to their soft integration. As described in Section \ref{sec:tools-nfv-sdn}, the Neutron service uses the Module Layer 2 (ML2) Plugin \cite{openstack} to provide networking services in a Cloud. The ML2 might control an ODL instance using the Neutron API, a REST API provided by ODL. However, the ONOS SDN Controller has gained space in both academia and industry and is currently the leading competitor of ODL. ONOS is the official distribution of the Open Network Foundation (ONF). Some industrial use case projects have been used ONOS, such as the Central Office Re-architected as a Datacenter (CORD), an NFV/SDN platform supported by major service providers (e.g., Google and Verizon). 

\begin{figure}[htbp]
	\centering
	\begin{adjustbox}{width=0.8\textwidth}
		\includegraphics[]{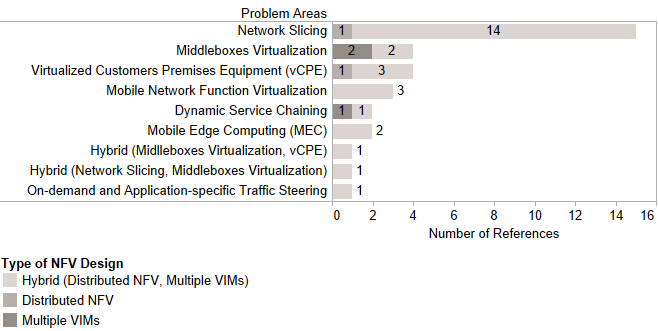}
	\end{adjustbox}
	\caption{Statistics related to the Problem Areas versus NFV Design.}
	\label{fig:problem-nfvdesign}
\end{figure}

The number of studies using OpenFlow 1.0 in NFV/SDN architectures has been steadily declining over the years (there is only one study in 2016) \cite{saridis_lightness:_2016}. As expected, there is a clear trend to use the most current OpenFlow version in recent studies \cite{poc_etsi_16,poc_etsi_26,poc_etsi_28,poc_etsi_34,8016637,7945849}. ForCES does not seem to have attracted the interest of the research community since only a few studies have used it as the southbound protocol \cite{haleplidis_forces_2014,haleplidis_towards_2014}.



\begin{figure}[htbp]
	\centering
	\begin{adjustbox}{width=0.8\textwidth}
		\includegraphics[]{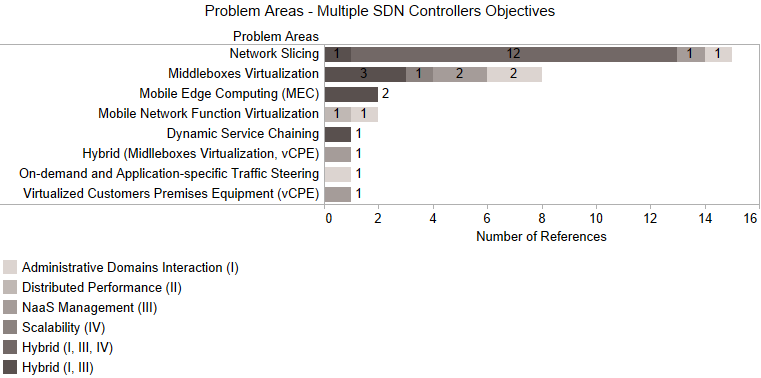}
	\end{adjustbox}
	\caption{Statistics related to the Problem Areas versus Objectives of Multiple SDN Controller.}
	\label{fig:problem-multisdncontfunc}
\end{figure}

In Section \ref{sec:nfvside}, the reader can observe that most solutions rely on both Distributed NFV and Multiple VIMs (see  Figure \ref{fig:problem-nfvdesign}), except for Wireless LAN and Wireless Mesh Networks. Particularly, 80\% of the studies for Network Slicing have used this type of NFV design, mainly because they are deployed under multiple administrative domains (see Figure \ref{fig:problem-multisdncontfunc}) including different types of network infrastructures (e.g., RAN, transport and core networks).

\begin{figure}[htbp]
	\centering
	\begin{adjustbox}{width=\textwidth}
		\includegraphics[]{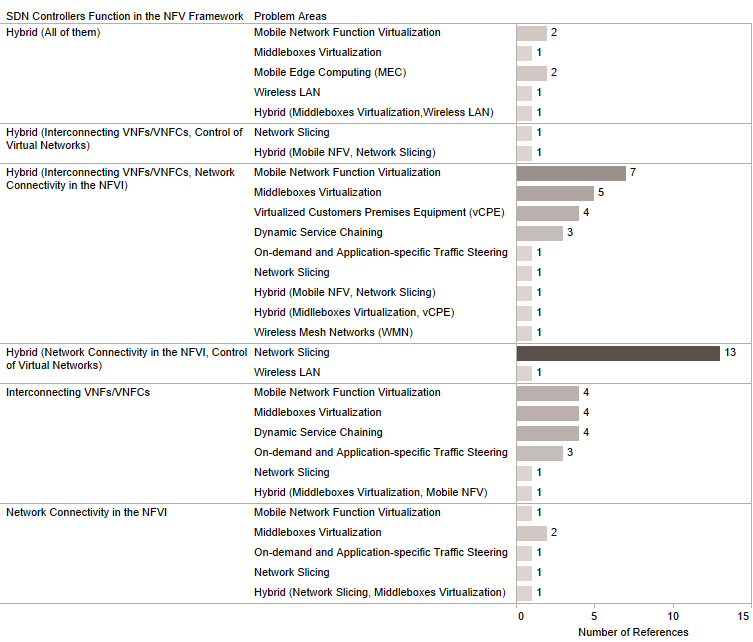}
	\end{adjustbox}
	\caption{Statistics related to the Problem Areas versus SDN Controller Function.}
	\label{fig:problem-sdncontfunc}
\end{figure}

As far as we are concerned to the SDN Controller Functions (see Figure \ref{fig:problem-sdncontfunc}), the primary focus for all areas was \textit{Interconnecting VNFs/VNFCs}, which can be considered indispensable for any NFV/SDN architectures. As SDN can deliver intelligent traffic steering, service chains can indeed benefit from this integration. Also, although the \textit{Control of Virtual Networks} function has been widely adopted in areas (e.g., Network Slicing with 75\% of studies), we have not identified its implementation in other scenarios such as On-demand and Application-specific Traffic Steering, Dynamic Service Chaining, and vCPE. We argue that the absence of this functionality is not an impediment to the implementation of solutions for traffic steering and dynamic SFC and it should be a functional requirement for vCPE. Please recall that vCPE must provide multi-tenant services, leaving the client responsible for the selection and configuration of their VNFs.

As shown in Figure \ref{fig:problem-multisdncontfunc}, the use of multiple SDN controllers has not been explored in Wireless LAN and Wireless Mesh Networks solutions, which may be a gap to be addressed for new NFV/SDN architectures with multiple controllers. Furthermore, we have identified a few studies dealing with \textit{Scalability} (elasticity mechanisms) and \textit{Reliability} (fault tolerance mechanisms) problems (see section \ref{sec:nfvside} and subsection \ref{subsec:sdnside_mul}), which require orchestrators to manage environments with multiple VIMs and SDN controllers in order to provide support to perform D-NFV management in several administrative domains or NFVI-PoPs. According to the 5GPPP, such characteristics are essential for the 5G network, since novel 5G technologies (e.g., Network Slicing) might impact the entire mobile network including mobile devices; radio access, transport, and core networks; and the cloud (local, regional, or global).  

Finally, the use of NFV/SDN architectures has been a growing trend for providing fast delivery of network services in a flexible and automated way for 5G networks. A certain NFV/SDN architecture creates an abstraction layer that unifies both computer and network resources and enables dynamic and application-specific traffic steering. Most studies have demonstrated a growing attention to the cost problem where they tried to reduce both CAPEX and OPEX costs. On the other hand, the use of NFV/SDN architectures for the Mobile Edge Computing (MEC) is incipient. Current solutions are still in their infancy, and better schemes are needed to provide distributed and dynamic SFC as well as to meet flow requirements. The problem of how to integrate MEC and network slicing in a unique NFV/SDN architecture has been slightly addressed \cite{akyildiz_wireless_2015,mwangama_towards_2015,nguyen_slicing_2014,medhat_multi-tenancy_2015,neves_selfnet_2016}.


\section{FUTURE RESEARCH DIRECTIONS: OPPORTUNITIES AND CHALLENGES}
\label{sec:challenges}

We have analyzed the selected studies and the surveys to have a clear view of the directions for future research efforts. We have identified some challenges in the design and implementation of NFV/SDN architectures. They are described in the following subsections.

\subsection{Deployment of Network Services} 

Service Function Chaining (SFC) provides an ordered list for service processing of traffic flows \cite{7350211}. The fast SFC deployment and provisioning of NFV and the centralized and flexible control of SDN have enabled new opportunities regarding this topic. However, solutions that provide better performance and optimal resource utilization in function deployment are still needed. Below, we describe some challenges related to NFV/SDN solutions regarding the deployment of network functions.

\begin{description}
\item[VNF Performance:] Virtualized network functions should meet the user's performance requirements, especially when the SDN Controller is a VNF. Current hypervisors must be optimized for fast packet processing in standard servers as a way to obtain high I/O speed, short transmission delays, and so on. Some initiatives are DPDK \cite{dpdk}, NetVM \cite{netvm}, and ClickOS \cite{clickos}. Further, the use of container-based virtualization, such as Docker containers, is of paramount importance in environments that require high-performance with low resource consumption such as edge devices (e.g., vCPEs, MEC, etc.) since they have relatively low capabilities compared to traditional NFV servers. However, there are some challenges to be considered when choosing the containers technologies for deploying VNFs in NFV/SDN architectures \cite{7945849,etsi_nfv_virt}:
	\begin{itemize}
	\item Orchestration: all containers share the same kernel as well as its services and configurations. This characteristic increases the complexity of orchestration and management platforms since the allocation process must take into account whether a VNF has unique needs in kernel, such as a particular module or configuration;
    \item Security: container-based virtualization has a broader attack surface than other virtualization techniques since its interface is more sophisticated than hardware emulation interfaces. Besides, it provides weaker functional isolation among instances since containers may require different kernel configurations that can conflict. It also offers more vulnerable performance isolation since containers in the same host can be placed under resource pressure (e.g., memory or CPU overconsumption) by external attacks or new instances.         
	\end{itemize}
 \item[VNFs Scheduling and Placement:] The scheduling and placement of VNFs impact the performance of Service Chaining significantly. For better performance, the physical resources should be used efficiently. Also, energy-efficient hardware and energy-aware network service placement remain some of the main challenges in NFV and solutions are still limited \cite{7243304}. Therefore, optimization and machine learning techniques are necessary to achieve optimal, automatic, and dynamic resource reservation, allocation, and migration of VNFs, considering a global view of the resources and the customer requirements. Integer programming and heuristic approaches can be used for VNFs Scheduling \cite{shen_vconductor:_2015} and Placement \cite{deng_vnguard:_2015}, considering resource constraints. Tools such as Google’s Borg \cite{borg}, Omega \cite{omega}, and Apache Mesos \cite{mesos} may be considered for scheduling of VNFs.
 \item[High-level Policies:] The definition of high-level policies is necessary to simplify the configuration of NFV Orchestrator operations, such as resource allocation and optimization mechanisms, and to meet the customers' requirements (interfaces to OSS/BSS). In this case, OpenStack's HOT (Heat Orchestration Template) and TOSCA (Topology and Orchestration Specification for Cloud Applications) template languages could be used  \cite{openstack}. 
\item[Traffic Steering:] In NFV/SDN solutions, traffic steering and network function deployment should be optimized jointly, providing a network-aware scheduling mechanism \cite{cerrato_toward_2015,lucrezia_introducing_2015} that deploys VNFs considering both the paths expressed in the forwarding graph and the network behavior (available bandwidth, latency, jitter, etc.). As a consequence, more variables are introduced, and heuristic algorithms should be created to reduce computing complexity.
\item[Elastic Network Function:] The dynamic service scaling at runtime provides better resource utilization, reducing both CAPEX and OPEX, and maintains service level requirements \cite{Szabo2015}. It is necessary that NFV/SDN solutions can scale (in/out or up/down) networking services and monitor both servers' and networks' resources to offer elastic, pay-as-used services.
\item[Orchestration:] Orchestration services are necessary for elastic, adaptable, and autonomic network function deployment, provisioning, and management. Tools such as OpenMANO \cite{openmano} and OpenBaton \cite{openbaton} might be used as a solution for NFV MANO (Management and Orchestration).
\end{description}

\subsection{Improving the Programmability} 

SDN and NFV are the critical enablers for realizing some of the expected features in 5G networks, such as network programmability, flexibility (e.g., network abstraction, infrastructure sharing, and reconfigurability), adaptability (e.g., self-healing, self-configuration, self-protection, and self-optimization) and capabilities (e.g., network slicing and MEC) \cite{5gppparchitecture}. However, some improvements should be provided so that the existing SDN standards such as OpenFlow can be applied in this type of scenario.  

OpenFlow is the most used protocol for the Southbound API in NFV/SDN solutions, as described in Section \ref{sec:sdnside}. However, currently, it does not support application layer packet processing. The application layer inspection and classification is necessary to provide fine-grained flow distribution for different network services, and thus to provide intelligent service chaining. 

Finally, OpenFlow is not suitable for Wireless Networks (e.g., WiFi, LTE, etc.) since flow tables just include rules for Ethernet-based switches. Wireless communication is more complex, as wireless links are time-varying and vulnerable to interference. For this, extensions must be implemented to allow WiFi programming rules, enabling the matching and monitoring of wireless frames \cite{schulz-zander_opensdwn:_2015}. Besides, SDN must provide support to Radio Access Network (RAN) virtualization infrastructures \cite{7926923}. In this case, SDN approaches must support both legacies (e.g., 3G and 4G) and new radio access technologies (e.g., 5G and narrowband Internet of Things, NB-IoT), ensuring radio resource isolation.
 
\subsection{Multi-Tenant, Multi-Service and Multi-Domain Support}

An NFV/SDN architecture that supports multiple domains or NFVI-PoPs is necessary for the provision of quality of service (QoS) and SLA enforcement in multi-tenant environments with end-to-end services. However, it remains a challenge since the orchestration functions must support the following features \cite{5gppparchitecture}:

\begin{itemize}
\item Multi-domain orchestration of diverse programmable infrastructure technologies (e.g., RAN, transport and core networks, data centers, etc.), possibly belonging to different operators;
\item Northbound interface for Network Slicing management, providing multi-tenancy and multi-service support;
\item End-to-end network slices that are flexible to the dynamic requirements of different services (e.g., IoT, smart cities, etc.) and mobile operators, providing a multi-service and context-aware adaptation of network functions;
\item Advanced autonomic network management platforms to address complexity in such scenarios.
\end{itemize}

Furthermore, studies are still being carried out to evaluate the impact of end-to-end slices on the RAN design. RAN Virtualization is currently under investigation and is one of the major obstacles to creating NFV/SDN architectures for 5G networks \cite{7926923}.

\subsection{Multiple SDN Controllers} 

NFV/SDN solutions could be used for the control and management of heterogeneous network resources (Optical, MPLS, IP, etc.), distributed in different geographical locations. Therefore, hierarchical and federated SDN Controllers must be used to meet scalability, availability, reliability, and end-to-end (multi-domain) provisioning requirements. Tools such as FlowVisor as well as North/East/WestBound interfaces from popular SDN Controllers (OpenDayLight \cite{opendaylight} and ONOS \cite{onos}) can be used to provide such solutions. 

\subsection{Security} 

In addition to current security problems that are unique to each technology \cite{7243304,Kreutz201414}, the NFV/SDN solutions also have security challenges related to the integration process, such as the lack of authentication and authorization mechanisms in the communication interfaces between SDN and NFV modules. Besides, by exploring network programmability, security services should be developed to deal with malfunctioning software (e.g., detecting and preventing exploits) or attacks caused by malicious adversaries (e.g., intrusion detection and prevention systems) such as Distributed Denial of Service (DDoS) \cite{DBLP:journals/corr/RomanLM16}. Such services should take into account attack surfaces at all levels of the infrastructure, including network, edge and core data centers, virtualization, and user devices. 

Furthermore, regarding 5G networks, security in network slicing is a complex task since there is resource sharing among slices and they may have different security policy requirements. This problem gets worse when we consider multi-domain scenarios. In this context, security solutions in the NFV/SDN architecture should provide mechanisms for resource isolation between slices, considering their impact on the entire infrastructure and providing security policy coordination among different domain infrastructures \cite{8039298}. 

\subsection{Extensibility and the Expressiveness of NFV/SDN Models} 

It is important to use a single model (framework) to address both NFV and SDN issues, instead of a combination that focuses on one problem at a time. This type of model eases implementation and the learning curve as well as reduces interdependencies (plugins to interconnect different frameworks) \cite{haleplidis_forces_2014}. 

\subsection{Standardization} 

Even with the existence of reference architectures defined by industry \cite{verizon}, an effort should be made towards standardization of an architecture that integrates the NFV and SDN technologies to simplify the work of researchers when providing new NFV/SDN solutions. This reference architecture must include standardized interfaces and resource catalogs \cite{shen_vconductor:_2015} so that new VNFs can be rapidly integrated and deployed into the system.

\section{THREATS TO VALIDITY}
\label{sec:threats}

This section describes several threats to the validity of this study, to evaluate the quality of this research. The potential validity threats to our study and the strategies for overcoming them are listed below:

\begin{description}
\item[Error-prone analysis:] The process of selection and extraction of studies were carried out by only one researcher (M. Bonfim), which may lead to subjectivity in the selection and inconsistencies in data extraction of the articles. To mitigate this threat, we have created a well-defined and extensive process to the studies selection and data extraction through the SLR protocol. Besides, two experienced researchers (K. Dias and S. Fernandes) checked and validated all the stages of the SLR (protocol, identification, and execution) and the summarization of results, to ensure the robustness and expressiveness of this work; 
\item[Data sources:] The primary studies were obtained from different web search engines and comprised of only academic studies. Even considering the Proof of Concepts of ETSI NFV ISG, this work is limited largely to academic expertise. It is important to conduct further research using industry data sources, such as the websites of companies, given that large companies (e.g., CISCO, Verizon, Juniper, HP, etc.) have invested heavily in NFV/SDN solutions. Furthermore, it is noteworthy that the search was done in April 2016, and the authors subsequently added some other bibliographies published after this period (until 2017), to provide a level of updating that allowed them to delimit some trends. Therefore, the volume of articles published in 2016 and 2017 may not reflect the current state nor indicate any trend for decrease in the interest of the research community;
\item[Meta-analysis:] The large number of selected studies in this SLR leads to a large variation in the reporting of NFV/SDN solutions regarding architecture and experimental techniques and dataset. This scenario has made our synthesis of the data largely qualitative because it has not been possible to carry out a meta-analysis to strengthen the differences among relevant studies. Conducting a statistical analysis of the data extracted from selected studies will need to be undertaken in the future.
\end{description}
\section{CONCLUSION}
\label{sec:conclusions}

Even with different purposes, NFV and SDN are complementary paradigms and technologies capable of providing one consolidated solution that offers the best of both technologies. NFV/SDN architectures are of paramount importance for a passage from the static design of conventional networks to an intelligent, open network environment. Therefore, this work proposed a Systematic Literature Review (SLR) for NFV/SDN architectures, intending to provide a profound understanding of such integrated designs. We aimed to identify the current trend in this field. For this, a total of 74 articles have been studied in-depth according to our predefined SLR protocol. Through comprehensive analysis and interpretation of the collected data, this SLR achieved three goals. First, we described the main characteristics (target environment and problems to solve) of integrated NFV/SDN solutions practices. Second, we compared their architecture designs (NFV framework design and tools, SDN APIs and place of SDN elements) and classified them against the presented taxonomy. Then, we discussed some opportunities and challenges for research work in the next generation of NFV/SDN architectures. 

\bibliographystyle{ACM-Reference-Format}
\bibliography{references}


\end{document}